\documentclass[
twocolumn,
amsmath,
amssymb,
aps,
prd,
superscriptaddress,
showpacs
]{revtex4-2}

\usepackage{graphicx}
\usepackage[colorlinks=true
,urlcolor=blue
,anchorcolor=blue
,citecolor=blue
,filecolor=blue
,linkcolor=blue
,menucolor=blue
,linktocpage=true
,pdfproducer=medialab
,pdfa=true
]{hyperref}
\usepackage[utf8]{inputenc}

\begin{document}
\title{A Convolutional Neural Network for Multiple Particle Identification in the MicroBooNE Liquid Argon Time Projection Chamber}

\newcommand{\Bern}{Universit{\"a}t Bern, Bern CH-3012, Switzerland}
\newcommand{\BNL}{Brookhaven National Laboratory (BNL), Upton, NY, 11973, USA}
\newcommand{\UCSB}{University of California, Santa Barbara, CA, 93106, USA}
\newcommand{\Cambridge}{University of Cambridge, Cambridge CB3 0HE, United Kingdom}
\newcommand{\StKates}{St. Catherine University, Saint Paul, MN 55105, USA}
\newcommand{\CIEMAT}{Centro de Investigaciones Energ\'{e}ticas, Medioambientales y Tecnol\'{o}gicas (CIEMAT), Madrid E-28040, Spain}
\newcommand{\Chicago}{University of Chicago, Chicago, IL, 60637, USA}
\newcommand{\Cincinnati}{University of Cincinnati, Cincinnati, OH, 45221, USA}
\newcommand{\CSU}{Colorado State University, Fort Collins, CO, 80523, USA}
\newcommand{\Columbia}{Columbia University, New York, NY, 10027, USA}
\newcommand{\FNAL}{Fermi National Accelerator Laboratory (FNAL), Batavia, IL 60510, USA}
\newcommand{\Granada}{Universidad de Granada, Granada E-18071, Spain}
\newcommand{\Harvard}{Harvard University, Cambridge, MA 02138, USA}
\newcommand{\IIT}{Illinois Institute of Technology (IIT), Chicago, IL 60616, USA}
\newcommand{\KSU}{Kansas State University (KSU), Manhattan, KS, 66506, USA}
\newcommand{\Lancaster}{Lancaster University, Lancaster LA1 4YW, United Kingdom}
\newcommand{\LANL}{Los Alamos National Laboratory (LANL), Los Alamos, NM, 87545, USA}
\newcommand{\Manchester}{The University of Manchester, Manchester M13 9PL, United Kingdom}
\newcommand{\MIT}{Massachusetts Institute of Technology (MIT), Cambridge, MA, 02139, USA}
\newcommand{\Michigan}{University of Michigan, Ann Arbor, MI, 48109, USA}
\newcommand{\Minnesota}{University of Minnesota, Minneapolis, MN, 55455, USA}
\newcommand{\NMSU}{New Mexico State University (NMSU), Las Cruces, NM, 88003, USA}
\newcommand{\Otterbein}{Otterbein University, Westerville, OH, 43081, USA}
\newcommand{\Oxford}{University of Oxford, Oxford OX1 3RH, United Kingdom}
\newcommand{\PNNL}{Pacific Northwest National Laboratory (PNNL), Richland, WA, 99352, USA}
\newcommand{\Pitt}{University of Pittsburgh, Pittsburgh, PA, 15260, USA}
\newcommand{\Rutgers}{Rutgers University, Piscataway, NJ, 08854, USA}
\newcommand{\StMarys}{Saint Mary's University of Minnesota, Winona, MN, 55987, USA}
\newcommand{\SLAC}{SLAC National Accelerator Laboratory, Menlo Park, CA, 94025, USA}
\newcommand{\SDSMT}{South Dakota School of Mines and Technology (SDSMT), Rapid City, SD, 57701, USA}
\newcommand{\Maine}{University of Southern Maine, Portland, ME, 04104, USA}
\newcommand{\Syracuse}{Syracuse University, Syracuse, NY, 13244, USA}
\newcommand{\TelAviv}{Tel Aviv University, Tel Aviv, Israel, 69978}
\newcommand{\Tennessee}{University of Tennessee, Knoxville, TN, 37996, USA}
\newcommand{\UTA}{University of Texas, Arlington, TX, 76019, USA}
\newcommand{\Tufts}{Tufts University, Medford, MA, 02155, USA}
\newcommand{\VTech}{Center for Neutrino Physics, Virginia Tech, Blacksburg, VA, 24061, USA}
\newcommand{\Warwick}{University of Warwick, Coventry CV4 7AL, United Kingdom}
\newcommand{\Yale}{Wright Laboratory, Department of Physics, Yale University, New Haven, CT, 06520, USA}

\affiliation{\Bern}
\affiliation{\BNL}
\affiliation{\UCSB}
\affiliation{\Cambridge}
\affiliation{\StKates}
\affiliation{\CIEMAT}
\affiliation{\Chicago}
\affiliation{\Cincinnati}
\affiliation{\CSU}
\affiliation{\Columbia}
\affiliation{\FNAL}
\affiliation{\Granada}
\affiliation{\Harvard}
\affiliation{\IIT}
\affiliation{\KSU}
\affiliation{\Lancaster}
\affiliation{\LANL}
\affiliation{\Manchester}
\affiliation{\MIT}
\affiliation{\Michigan}
\affiliation{\Minnesota}
\affiliation{\NMSU}
\affiliation{\Otterbein}
\affiliation{\Oxford}
\affiliation{\PNNL}
\affiliation{\Pitt}
\affiliation{\Rutgers}
\affiliation{\StMarys}
\affiliation{\SLAC}
\affiliation{\SDSMT}
\affiliation{\Maine}
\affiliation{\Syracuse}
\affiliation{\TelAviv}
\affiliation{\Tennessee}
\affiliation{\UTA}
\affiliation{\Tufts}
\affiliation{\VTech}
\affiliation{\Warwick}
\affiliation{\Yale}

\author{P.~Abratenko} \affiliation{\Tufts} 
\author{M.~Alrashed} \affiliation{\KSU}
\author{R.~An} \affiliation{\IIT}
\author{J.~Anthony} \affiliation{\Cambridge}
\author{J.~Asaadi} \affiliation{\UTA}
\author{A.~Ashkenazi} \affiliation{\MIT}
\author{S.~Balasubramanian} \affiliation{\Yale}
\author{B.~Baller} \affiliation{\FNAL}
\author{C.~Barnes} \affiliation{\Michigan}
\author{G.~Barr} \affiliation{\Oxford}
% ...

\collaboration{The MicroBooNE Collaboration} 
\thanks{microboone\_info@fnal.gov}\noaffiliation

\begin{abstract}
We present the multiple particle identification~(MPID) network, a convolutional neural network~(CNN) for multiple object classification, developed by MicroBooNE.  MPID provides the probabilities that an interaction includes an $e^-$, $\gamma$, $\mu^-$, $\pi^\pm$, and protons in a liquid argon time projection chamber~(LArTPC) single readout plane. The network extends the single particle identification network previously developed by MicroBooNE~\cite{ub_singlePID}. MPID takes as input an image either cropped around a reconstructed interaction vertex or containing only activity connected to a reconstructed vertex, therefore relieving the tool from inefficiencies in vertex finding and particle clustering. The network serves as an important component in MicroBooNE's deep learning based $\nu_e$ search analysis. In this paper, we present the network's design, training, and performance on simulation and data from the MicroBooNE detector. 
\end{abstract}

\maketitle

\section{Introduction}
\label{sec:intro_general}
A series of liquid argon time projection chamber~(LArTPC) detectors have been or are being deployed at Fermilab as part of the Short-Baseline Neutrino~(SBN) program~\cite{sbn_proposal} along the Booster Neutrino Beamline~(BNB~\cite{BNB}) and as part of the long-baseline program of the Deep Underground Neutrino Experiment~(DUNE)~\cite{dune_proposal}. 
The MicroBooNE experiment~\cite{ub_detector}, part of the Fermilab SBN program, has been operating since 2015, collecting data accumulated during beam-on and beam-off time periods.  

 
MicroBooNE operates a 170 ton~(85 ton active) LArTPC placed 470~m from the BNB target at Fermilab. The LArTPC is 10.4~m long, 2.6~m wide and 2.3~m high. 
The detector has three readout wire planes with 2400 readout wires on the two induction planes and 3456 readout wires on the collection plane~\cite{wirepaper}.  
Wires are installed with two induction planes oriented at $\pm60^{\circ}$ with respect to the vertical collection plane at a wire pitch of 3~mm.  
An array of 32 PMTs are installed behind the collection plane to detect the scintillation light from argon ionization caused by charged final state particles from neutrino interactions~\cite{pmtpaper}. 
The TPC readout time window is 4.8~ms and is digitized into 9600 readout time ticks. 
Charged particles in liquid argon produce ionization electrons, which drift to the readout wire planes in an electric field of 273~V/cm. 
It takes 2.3~ms for an ionization electron to drift across the full width of the detector.  

The MicroBooNE LArTPC continuously records charge drifted and its arrival time on each wire. 
A software trigger, based on PMT signals, records an event triggered by the BNB beam spill if the interaction light detected by the PMT array is above a set threshold. 
Each event consists of data collected from 1.6~ms before the trigger and 3.2 ms after the trigger.  
Therefore, each event has three sets of TPC data for each wire on all three planes.  
A truncation of the wire readout is performed around the trigger results so that the two induction planes have resolutions of 2400 wires~$\times$ 6048 readout ticks, while the collection plane has a resolution of 3456 wires~$\times$6048 readout ticks. 
Wire and time data can be converted into an image format (charge on each wire versus drift time) using the software toolkits LArSoft~\cite{uboonecode} and LArCV~\cite{larcv} while maintaining high resolution in wire, time and charge amplitude space. These information-rich LArTPC images are suitable for applying deep learning tools. 
In consideration of computing resources, images for deep learning tools are compressed along the time tick axis by a factor of six. Pixel values are merged by a simple sum. Images become 2400 wires~$\times$~1008~ticks and 3456 wires~$\times$~1008 ticks for the induction and collection planes, respectively.  
This corresponds to an effective position resolution of 3.3~mm~\cite{e_velocity} and 3~mm~\cite{wirepaper} along the time tick and wire number directions, respectively.  


Convolutional neural networks~(CNN), deep learning networks commonly applied to image processing applications, are currently used across neutrino and high energy physics experiments~\cite{dl_nature}.  
For accelerator neutrino experiments, NOvA has applied a CNN as a neutrino event classifier~\cite{nova_cvn} in its $\nu_\mu\to\nu_e$ oscillation measurement~\cite{nova_osc, nova_osc_new} and its neutral-current~(NC) coherent $\pi^0$ production measurement~\cite{nova_pi}.  
NOvA has also demonstrated a context-enriched particle identification network~\cite{nova_context}.  
MINERvA has developed CNN tools to determine neutrino interaction vertices and study possible biases due to models used in the large simulated training sample~\cite{minerva}.
The NEXT experiment has also used a CNN classifier to perform particle content studies at candidate neutrinoless double beta decay vertices~\cite{next}.  

A variety of deep learning techniques have been used in neutrino LArTPC experiments.  
In MicroBooNE, a CNN for assigning probabilities of particle identities for single particles in the MicroBooNE LArTPC has been demonstrated on simulated data in Ref.~\cite{ub_singlePID}.  
A semantic segmentation network for LArTPC data~\cite{ssnet, microboonecollaboration2020semantic} has been used for $\pi^0$ event reconstruction~\cite{ub_ccpi0}, vertex finding, and track reconstruction~\cite{vtxfinding}.  
The DUNE experiment has recently presented an updated long-baseline neutrino oscillation sensitivity study incorporating a CNN for neutrino event selection and background rejection~\cite{dune_cnn}.

In this article, we present our study in developing and applying a multiple particle identification~(MPID) network with the task of multiple binary logistic regression problem solving in MicroBooNE.  
It is the first demonstration of the performance of a CNN on LArTPC data including systematic uncertainties, and the first particle identification network applied to LArTPC datasets.  
The MPID network extends the functionality of MicroBooNE's previously-described single PID CNN network~\cite{ub_singlePID}.  
It does not require pre-processing of image data to identify and filter selected pixels in an image assumed to be produced by a specific particle.   
The network provides simultaneous prediction scores for particle existence probabilities in the same image among five different particle species: electrons ($e^-$), photons~($\gamma$), muons ($\mu^-$), charged pions ($\pi^\pm$) and protons ({\it p}). 
The network is a particularly useful tool for data analysis of particle interactions in LArTPC detectors, since the region of an interaction vertex often contains many particles.  

The MPID algorithm can take as input a LArTPC image with a fixed 512$\times$512 pixel scale. A detailed description of the network design and training for MPID is given in Section~\ref{sec:intro_cnn}.  
When used in MicroBooNE's deep learning based low-energy excess $\nu_e$ (LEE $1e${\small-}$1p$) search analysis, the MPID network is primarily applied to images that contain candidate reconstructed neutrino interaction vertices as well as all reconstructed topologically connected activity.  
MPID predictions are derived based on the full information of all energy depositions topologically connected to the vertex, particularly the first few centimeters of final-state particles' trajectories, which are critical for particle identification.  
In the $\nu_e$ search, the network is also applied to more inclusive images roughly cropped around the interaction vertex.  This is a new feature compared with the single PID network, which takes as input only images containing filtered, reconstructed hits. 
Cropping around the interaction vertex allows re-evaluation of charge missing from the former topologically-connected image, but is nonetheless present near the vertex, such as photon showers from final-state $\pi^0$s.  
This feature of the MPID network can help MicroBooNE suppress important photon backgrounds to a LEE search, as observed by MiniBooNE~\cite{miniboone}. 
We demonstrate this feature's robustness against the presence of LArTPC activity such as cosmic ray tracks that are uncorrelated with signal features of interest.

In this paper, we are not prepared to show full performance in the context of a physics analysis, but we can present some specific measures of network performance. Section~\ref{sec:mc} shows the efficiency of the different particle scores on idealized events containing $e^-$, $\mu^-$, and {\it p}; Section~\ref{sec:data} shows data-simulation agreement on samples highly enriched in certain signal topologies; and Section~\ref{sec:nue_mc} shows efficiency and background rejection performance for $\nu_e$ and some specific backgrounds.

\section{Multiple particle convolutional neural network}
\label{sec:intro_cnn}

\subsection{Network design}

\begin{figure}[htb!pb]
\centering
\includegraphics[width=6.5cm]{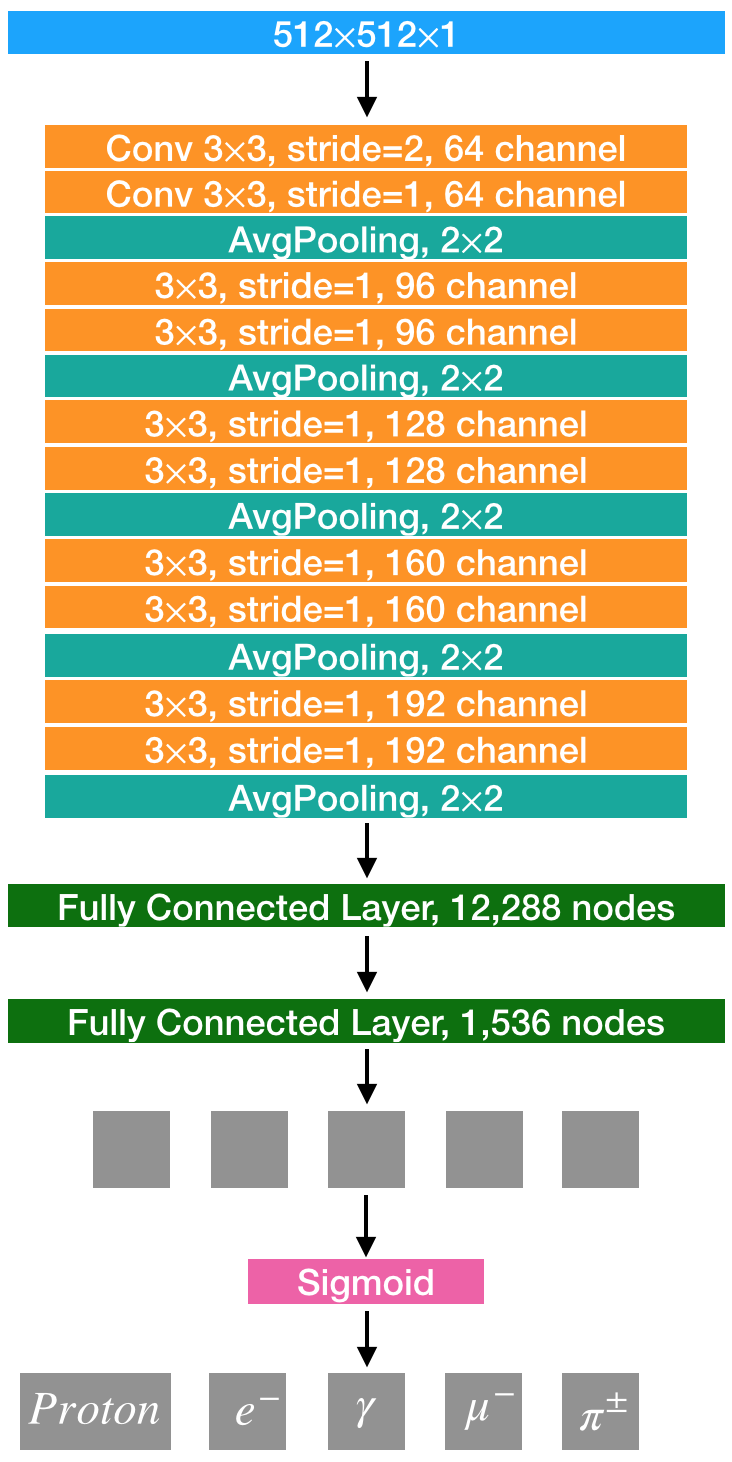}
\caption{MPID network scheme. The output has five numbers. Each of the values is between 0 and 1, representing the probabilities of corresponding particles in the given LArTPC image. \label{fig:mpid_design}}
\end{figure}

The MPID network applies a typical CNN~\cite{cnn} structure for the task of multiple object classification, which is summarized in block diagram form in Fig.~\ref{fig:mpid_design}.  
Input images have a resolution of 512 $\times$ 512~(1.5~m$\times$1.5~m) pixels, which generally matches the size of neutrino-induced activity in MicroBooNE. 
A series of ten convolutional layers are applied to the image for extracting high-level features. 

The first convolutional layer has a stride~(shift unit of the convolution calculation) of two with the goal of reducing the LArTPC images' sparsity and increasing feature abundance at the beginning of the algorithm. 
Following convolutional layers have a stride of one, a block of two convolution layers with a kernel size of three, followed by a pooling layer that is repeated five times.  
An average pooling layer is applied at every other convolutional layer to contract the spatial dimension. Then following the pooling layer is a rectifier activation function~(ReLU)~\cite{relu} for adding non-linearities to the network, as well as a group normalization operator~\cite{GN} to avoid early overfitting. 

Two fully connected layers with 192$\times$8$\times$8 nodes and 192$\times$8 nodes are applied to combine the features derived by convolutional layers.  
Output of the fully connected layers is a vector with five floating point numbers, each representing a confidence score for a target particle type to be present in an image. The score is interpreted as a normalized probability after applying a sigmoid function~\cite{pytorch_sigmoid}. The algorithm is optimized by minimizing the sum of binary cross entropy loss~\cite{pytorch_bce} across target particle types.   
In this way the prediction categories are not exclusive between particles.

\begin{figure}[htbp!]
\centering
\includegraphics[width=8.6cm]{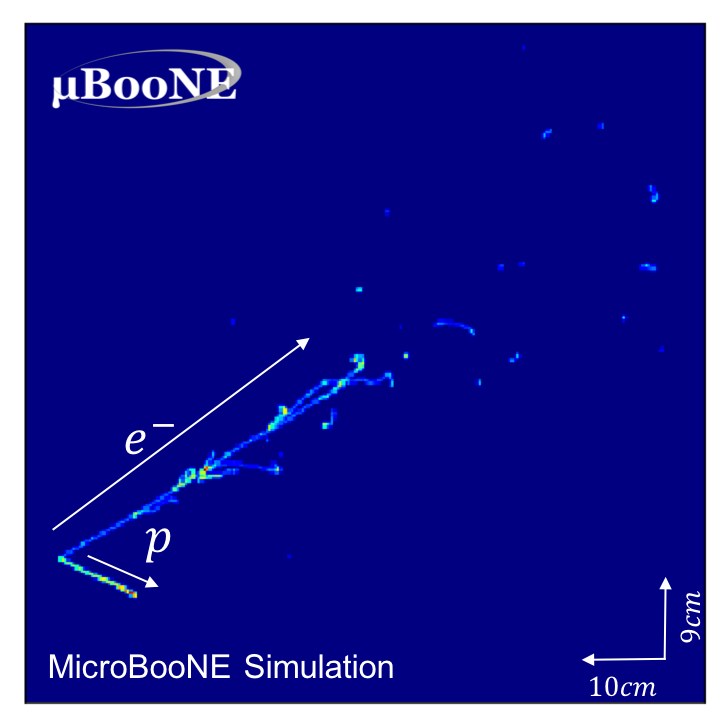}
    \begin{tabular}{cccccc}
    \hline
     &  {\it p} & $e^-$ & $\gamma$ & $\mu^-$ & $\pi^\pm$ \\
    \hline
    MPID Score & 0.99 & 0.98 & 0.06 & 0.01 & 0.02 \\
    \hline
    \end{tabular}
\caption{MPID example of an $1e${\small-}$1p$ topology with a tabulated output of particle scores.  This image is generated by concatenating a {\it p} and an $e^-$ at the same vertex.  Scores indicate high probabilities of having a {\it p} and $e^-$ in the image. The image applied to MPID has 512~$\times$~512 pixels. A zoom-in image of 250~$\times$~250 pixels is shown for visualization.\label{fig:mpid-1e1p-exp}}
\end{figure}

Figure~\ref{fig:mpid-1e1p-exp} shows one example of the input and output of the MPID network during inference.  
In this case, the input image has one $e^-$ and one {\it p} concatenated at the same vertex, a typical signal interaction topology for an interaction-channel-exclusive $1e${\small-}$1p$ search, as implemented in the MicroBooNE deep learning based LEE analysis.  
The MPID network calculates as output the five floating point numbers described in the previous paragraph, or ``particle scores,'' that correspond to the inferred probability to have each type of particle present in the image. In this example, high scores of 0.99 and 0.98 are given for {\it p} and $e^-$ in the image and low scores of 0.06, 0.01 and 0.02 are provided for $\gamma$, $\mu^-$ and $\pi^\pm$.  

\begin{figure}[htpb!]
\centering

    \includegraphics[width=8.6cm]{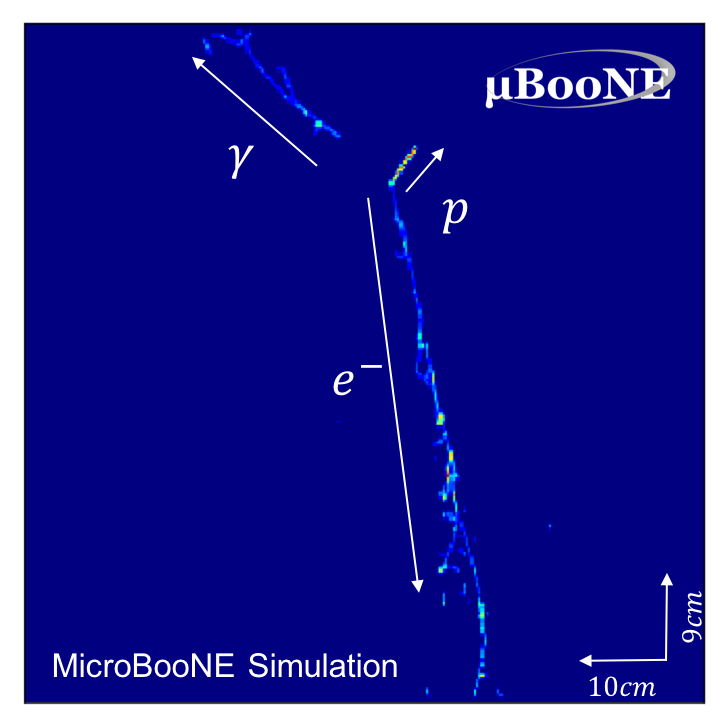}
    \begin{tabular}{cccccc}
    \hline
     &  {\it p} & $e^-$ & $\gamma$ & $\mu^-$ & $\pi^\pm$ \\
    \hline
    MPID Score & 0.89 & 0.95 & 0.85 & 0.06 & 0.17 \\
    \hline
    \end{tabular}

\caption{MPID example of an $1e${\small-}$1\gamma${\small-}$1p$ topology with a tabulated output of particle scores.  This image is generated by concatenating three particles at the same vertex.  Scores indicate higher probabilities of having {\it p}, $e^-$ and $\gamma$ in the image. The image applied to MPID has 512~$\times$~512 pixels. A zoom-in image of 250~$\times$~250 pixels is shown for visualization. \label{fig:mpid-1g1p-exp}}
\end{figure}

Figure~\ref{fig:mpid-1g1p-exp} shows another example of the input and output for the network during inference. The input image has one $\gamma$, one $e^-$ and one {\it p} produced at same vertex, which would in principle be rejected in an exclusive $1e${\small-}$1p$ search.  
Again, the MPID calculates scores that correspond to containing each particle in the image.  
High scores of 0.89, 0.95 and 0.85 are for {\it p}, $e^-$ and $\gamma$ in the image and low scores of 0.02 and 0.08 are found for $\mu^-$ and $\pi^\pm$ in the image.  
We also note for total clarity that the photon particle score is indicative not of the predicted \emph{total number} of photons in the image, but rather the probability that \emph{any} photons are present in the image.  
The former judgement, as well as the capability to identify the particle content of specific sub-features within an image, is not within the scope of the MPID algorithm.  

\subsection{Training and Test Samples}
\label{training_sample}
Training and test samples for the MPID CNN are produced with a customized event generator that uses LArSoft~\cite{uboonecode} and LArCV~\cite{larcv}.  
Detector processes are simulated with the GEANT4~\cite{geant4_1,geant4_2,geant4_3} simulation tool. 

The first generator step produces a 3D vertex uniformly distributed in the MicroBooNE LArTPC. 
The second step generates a random number of particles from $e^-$, $\gamma$, $\mu^-$, $\pi^\pm$, and $p$ options.  
All particles are generated at the vertex from the first step with isotropic directions.  
The multiplicities for the total number of particles allowed in each image are randomly distributed  between two and four.  
The multiplicity for each particle type is allowed to vary randomly between zero and two.  
Such a configuration will include as a subset final-state interaction vertex topologies that we are searching for or trying to reject in MicroBooNE analyses, such as $1e${\small-}$1p$, $1\mu${\small-}$1p$ and $1\gamma-1p$, as well as non-signal ones, such as $2\mu$ or $2e$.  
This generation strategy purposefully does not rely on any of the standard neutrino final-state generators~\cite{genie} to avoid possible biasing the MPID network via inclusion of possibly-incorrect kinematic or multiplicity information provided by the generator.  
Moreover, this training model will produce a more robust particle identification tool capable of producing unbiased results for a much broader range of vertex-generating physics processes.  
Finally, high multiplicity topologies generated in this randomized training samples help the network to activate more nodes and learn more parameters for classification.  

Each particle is generated with a single particle simulation package, where no neutrino interaction model kinematics are assumed.  
For 80\% of the training and test samples, particles are simulated with kinetic energies between 60~MeV and 400~MeV for protons and between 30~MeV and 1~GeV for other particles.  
For the other 20\% of the training and test samples, particles are simulated with kinetic energies between 40~MeV and 100~MeV for protons and between 30~MeV and 100~MeV for other particles. Particles are generated with a flat energy distribution.
Energy ranges are chosen based on the BNB neutrino energy distribution and the analysis priority towards the lower energy range.
We generated 60,000 simulated events for training and 20,000 images for testing.  
The images are intentionally generated without overlaying cosmic rays on simulated images to retain separation capabilities for $\mu^-$.  
Images used for training, testing and inference are from the better performing collection plane only~\cite{Adams_2020}, similar to networks described in Ref.~\cite{ub_singlePID} and Ref.~\cite{ssnet}.  
This choice serves to reduce the network's reliance on upstream reconstruction steps, such as the matching of pixels from different wire planes.  

\subsection{Network Training}

The loss of the network is defined using the BCEWithLogitsLoss~\cite{pytorch_bce} function in PyTorch taking the output layer~(five floating point number) as input. The BCEWithLogitsLoss function combines a sigmoid~\cite{pytorch_sigmoid} operator with the binary cross entropy calculation.  
During training, we applied an initial learning rate of 0.001.  
Batch sizes of 32 and 16 are chosen for the training and test processing. Training is processed with one single NVIDIA 1080 Ti graphics card. Regularization methods of dropout~\cite{dropout} and group normalization~\cite{GN} are applied to avoid early overfitting during training. 

\begin{figure}[htb!pb]
\centering
\includegraphics[width=8.6cm]{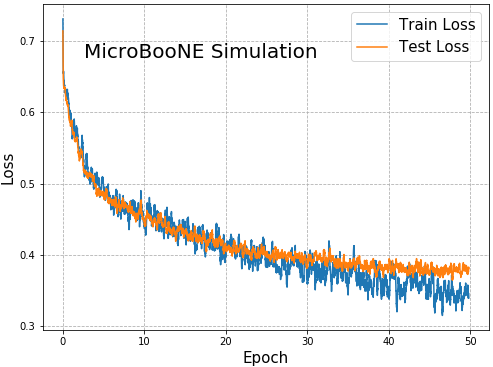}
\includegraphics[width=8.6cm]{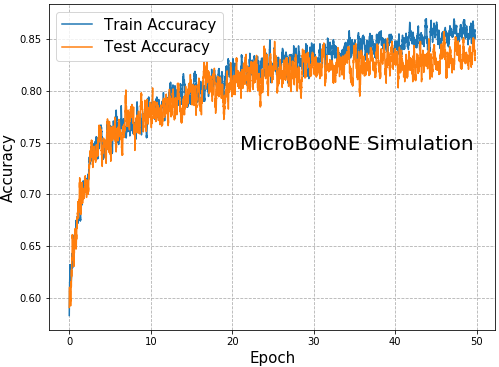}
\caption[]{Losses of training and test events during training~(top). Accuracies of training and test events during training~(bottom).}
\label{fig:mpid_training}
\end{figure}

\begin{figure*}[htb!pb]
\centering
\includegraphics[width=7.38cm]{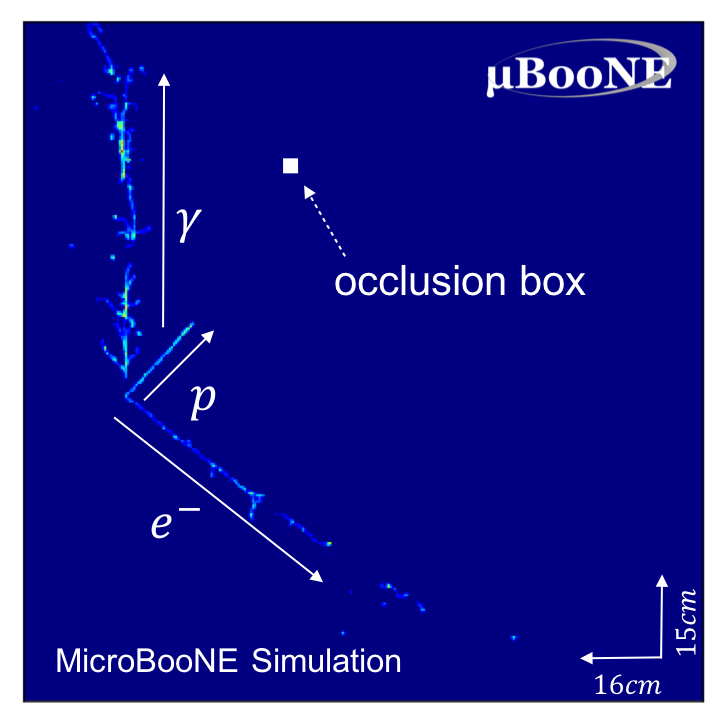}
\par 
\includegraphics[width=8.6cm]{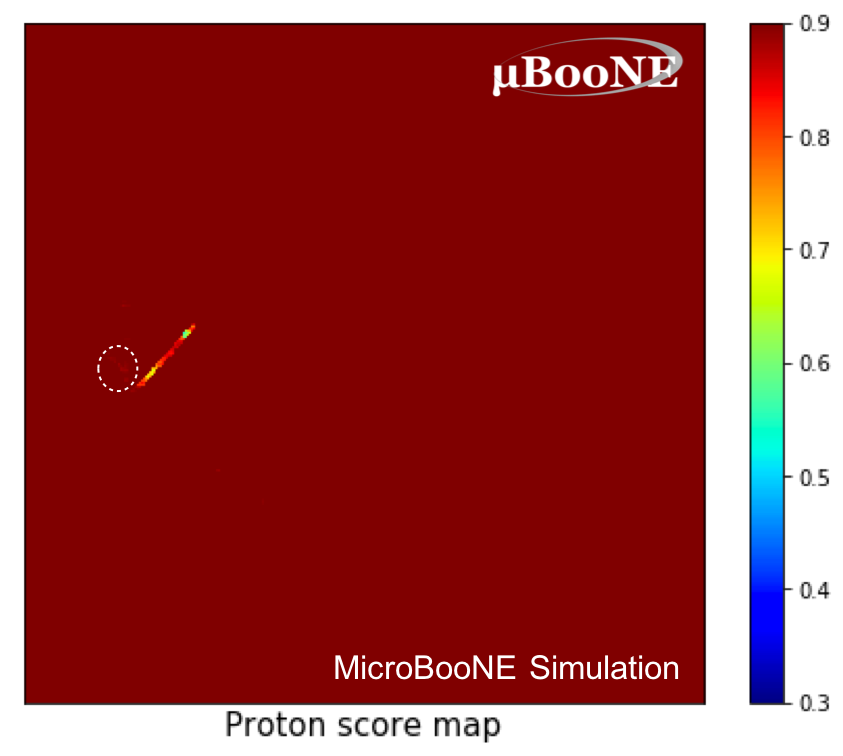}
\includegraphics[width=8.6cm]{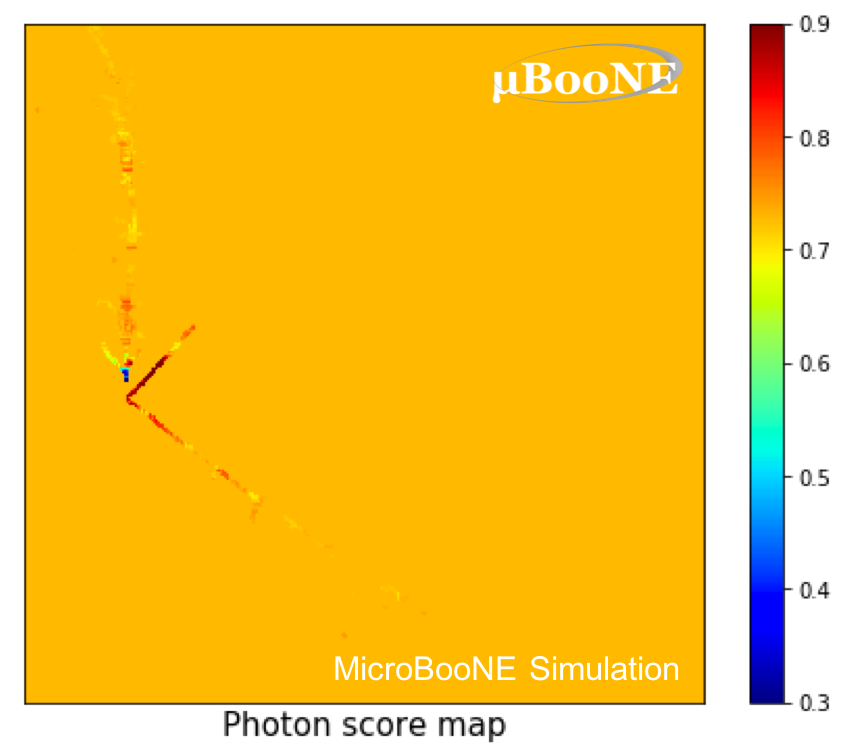}
\caption[]{Simulated $1e${\small-}1$\gamma${\small-}1{\it p} final state event example~(top). {\it p} score map~(bottom left), {\it p} scores decrease as occluded region crosses the {\it p} pixels. $\gamma$ score map~(bottom right), $\gamma$ scores decrease as pixels in the trunk region of the $gamma$ shower are occluded and increase as the trunk region of the e- shower are occluded.}
\label{fig:occlusion}
\end{figure*}

An accuracy is calculated while the training is monitored for loss. Accuracy is defined as the fraction of predicted labels matching the truth labels with a threshold value of 0.5 per event. MPID training curves of accuracies and losses are shown in Fig.~\ref{fig:mpid_training}.  
After epoch 29, the test curve continues to improve but does not keep up with the training curve.
With the consideration of not introducing bias from the training sample,
 we checked weights around epoch 29 and selected the one with best accuracy on the test sample.

\subsection{MPID Occlusion Analysis}

We applied an occlusion analysis~\cite{occlusion}
to determine whether the MPID network has calculated its predictions using image features associated with underlying physics for example, dE/dx at the first pixels of a particle~(referred as the trunk region of a particle), as opposed to other extraneous features in the image.  
The strategy is to feed the network an image partially masked to check how the MPID responds to the masked image.  
The occlusion analysis places a 9$\times$9 pixel box in the top left corner of the image, which masks all pixels in the occlusion box with zero values.
With this box placed, we then apply the MPID network to the masked image and plot at that center pixel the produced score value.  
This process is then repeated for each pixel as the occlusion box scans along the x and y axis of the image.  
Figure~\ref{fig:occlusion} shows an example of the occlusion box placed on the image. 
After scanning the whole image, we obtain score maps showing the MPID responses to each occlusion box placement location.  
A lowered score for a particular pixel in occlusion images indicates that the masked region contains topological information valuable for determining the identity of that particular particle.  

A simulated interaction image with one $e^-$, one $\gamma$ and one {\it p} at the same vertex, shown in Fig.~\ref{fig:occlusion} is chosen to demonstrate the occlusion analysis.  
The bottom left panel in Fig.~\ref{fig:occlusion} shows the {\it p} scores from the occlusion study on the input image.  
The {\it p} score drops significantly as the proton track's Bragg peak region, where strong {\it p} dQ/dx features exist, is masked.  
This indicates that the MPID network is properly identifying and leveraging features associated with the {\it p}'s unique energy deposition density profile.  
It can also be seen that a few pixels in the circle with very high pixel values in the pictured shower are mildly misinterpreted as {\it p}-like features.

The bottom right panel in Fig.~\ref{fig:occlusion} shows the $\gamma$ scores from occlusion analysis of the same input image.  
From the occlusion image, it is clear that a few key physics features of $\gamma$-containing images have been properly learned by the MPID network.  
There are two critical features in the particle trunk region for $e^-/\gamma$ separation: the projected trunk region dE/dx difference and the presence or absence of a gap between the trunk and the interaction vertex.
One can see the $\gamma$ score drops to near 0.3 when the trunk region of the $\gamma$ (rather than the gap region between $\gamma$ and vertex) is masked.  
We also applied an occlusion analysis to images with single $\gamma$ images to confirm that $\gamma$ scores drop and $e^-$ scores increase as the $\gamma$ trunk region is masked.
We observe in this example that the $\gamma$ score also increases to near one when nearby pixels connecting the {\it p} and $e^-$ are masked, since this produces more gaps between different particles. 
The $e^-$ score does not change much as we move the occlusion box around since there are overwhelming $e^-$-like features in the image from both the $e^-$ and $\gamma$. This observation indicates that consideration of both $e^-$ and $\gamma$ scores is likely important in attaining good $e^-$ and $\gamma$ separation with the MPID network.

\section{Performance on Simulation}
\label{sec:mc}

To provide a first look at the capabilities of the trained MPID algorithm, we present particle score results returned from the test images generated using the same method applied in producing training images.  
This section is divided into discussions of individual final state vertex and particle topologies of interest to MicroBooNE physics analyses, with occasional reference to a larger set of complimentary final state particle combinations located in Appendix~\ref{app1}.  

We primarily focus on two generated test samples with particles $1\mu${\small-}$1p$ and $1e${\small-}$1p$ in the final state, which are not used in training.  10,000 events are generated in each sample. These samples are generated with the same customized event generator described in Section~\ref{training_sample}. 
Vertices are uniform in the detector with one proton and one corresponding lepton. Kinetic energies of the protons are between 50~MeV and 400~MeV, while kinetic energies of leptons are generated between 50~MeV and 1~GeV.  
  The $1e${\small-}$1p$ final state dataset has a similar final state as the target events of MicroBooNE's deep learning based LEE $1e${\small-}$1p$ analysis.  
  The $1\mu${\small-}$1p$ dataset has a final state similar to a MicroBooNE $\nu_{\mu}$ selection analysis, described in Section~\ref{sec:1u1p}, that will be used to constrain the beam-intrinsic backgrounds in the LEE search. 
  For complimentary final state particle combinations located in Appendix~\ref{app1}, generated protons, muons, and electrons are generated with similar requirements as given above, while pions and gammas follow requirements similar to those of muons and electrons, respectively.  
  For completeness, Appendix~\ref{app1} includes descriptions of MPID performance all combinations of the five considered final state particle types, excepting the $1\mu${\small-}$1p$ and $1e${\small-}$1p$ sets described in this section.  

\subsection{$1\mu${\small-}$1p$ Simulated Sample}

\begin{figure}[htb!pb]
\centering
\includegraphics[width=8.6cm]{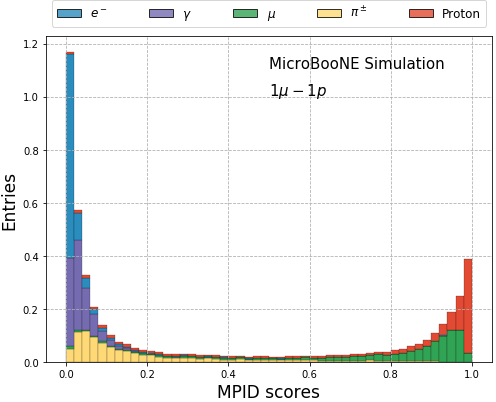}
\caption{~MPID score distributions for the probabilities of {\it p}, $\mu^-$, $e^-$,$\gamma$, $\pi^\pm$ on the $1\mu${\small-}$1p$ validation sample. \label{fig:mpid-val-1mu1p}}
\end{figure}

Figure~\ref{fig:mpid-val-1mu1p} shows stacked MPID scores of five particle hypothesis for the $1\mu${\small-}$1p$ simulated test dataset. A similar plot showing a complementary inverted final-state configuration~($Ne${\small-}$N\gamma${\small-}$0\mu${\small-}$N\pi${\small-}$0p$) is shown in Fig.~\ref{fig:PS_14} in appendix~\ref{app1}. One can see between Fig.~\ref{fig:mpid-val-1mu1p} and Fig.~\ref{fig:PS_14} the MPID network provides good separation between track-like and shower-like particles with {\it p} and $\mu^-$ scores concentrated near one and $e^-$ and $\gamma$ piled up near zero and vice versa in the complementary sample.

\begin{figure}[htb!pb]
\centering
\includegraphics[width=8.6cm]{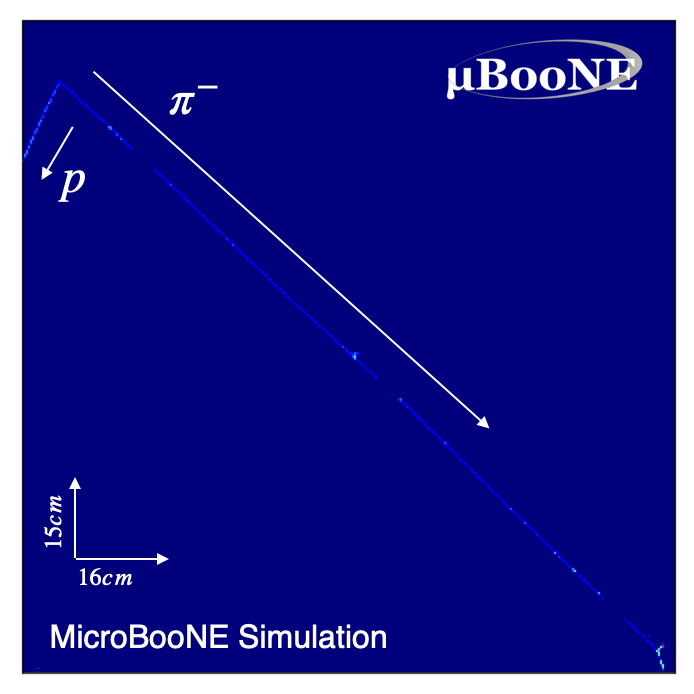}
\vspace{1pt}
\includegraphics[width=8.6cm]{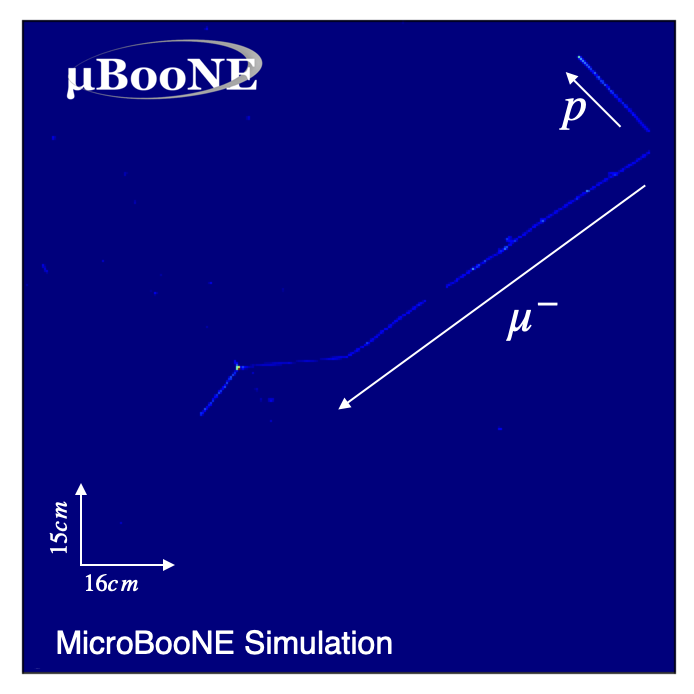}
\caption{ Simulated $1\pi${\small-}$1p$~(top) and $1\mu${\small-}$1p$~(bottom) events. MPID predicts a high $\mu^-$ score at 0.93 and a low $\pi^\pm$ at 0.10 for the $1\pi^-${\small-}$1p$ event where no kink is present (top). MPID predicts a high $\pi^\pm$ score at 0.97 and a low $\mu^-$ score at 0.27 for an $1\mu${\small-}$1p$ event (low) where the muon scatters and has a kink on its track trajectory. \label{fig:pi_mu_score}}
\end{figure}

The plot also shows a good separation between $\mu^-$ and $\pi^\pm$ using MPID, with a low score distribution for $\pi^\pm$. Separation between $\mu^-$ and $\pi^\pm$ comes from the fact that $\pi^\pm$ have higher rates of nuclear scattering than the $\mu$, and the $\pi^\pm$ can have a kink point where they decay as noted in Ref.~\cite{ub_singlePID}. The network is likely keying primarily off of visible kinks in a particle's trajectory in order to identify $\pi^\pm$ and the absence of visible kinks in a particle trajectory to identify $\mu^-$. By checking MPID over a hand scanning of images from a $1\pi^\pm${\small-}$1p$ sample, we notice MPID predicts high $\pi^\pm$ score and low $\mu^-$ score when the kink is visible, and vice versa when the kink is not visible. Fig.~\ref{fig:pi_mu_score} shows examples of predicting a high $\mu^-$ score for an $1\pi^\pm${\small-}$1p$ event where no kink is present and and predicting a high $\pi^\pm$ score for an $1\mu${\small-}$1p$ event where the muon scatters and has a kink on its track trajectory.

\begin{figure}[htb!pb]
\centering
\includegraphics[width=8.6cm]{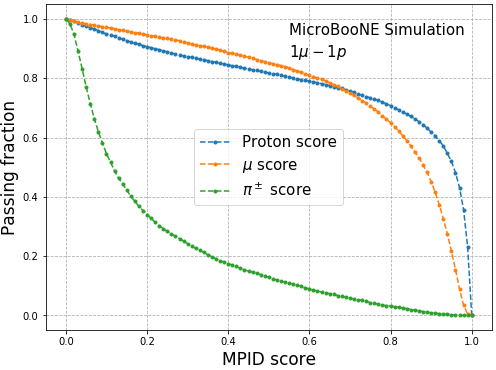}
\vspace{1pt}
\includegraphics[width=8.6cm]{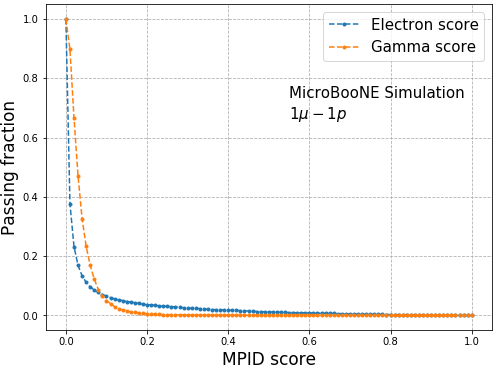}

\caption[]{MPID passing fractions for track-like particles~(top) of {\it p}, $\mu^-$ and $\pi^\pm$ on the $1\mu${\small-}$1p$ validation sample. MPID passing fractions for shower-like particles~(bottom) of $e^-$ and $\gamma$ on the $1\mu${\small-}$1p$ validation sample.}
\label{fig:mpid-effi-1mu1p}
\end{figure}


To perform particle identification as part of a neutrino event selection analysis, a set of selections are usually applied to particle score variables; these cuts will have associated impacts on total signal selection efficiencies.
Figure~\ref{fig:mpid-effi-1mu1p} shows the passing fractions for track-like particles in the $1\mu${\small-}$1p$ dataset. Similar plots of the complementary configuration~($Ne${\small-}$N\gamma${\small-}$0\mu${\small-}$N\pi${\small-}$0p$) are shown in Fig.~\ref{fig:PS_14} in appendix~\ref{app1}.
Passing fraction is defined as the percentage of events with an MPID particle score above a specified value; a tested set of events will have a passing fraction calculated for each particle type.  
The cut value for each particle score is varied between 0 and 1 with a step size of 0.01. 
For example the blue dotted line shows the passing fraction of {\it p} in the image at each {\it p} score cut value.
Figure~\ref{fig:mpid-effi-1mu1p} also shows the passing fractions for shower-like particles in the $1\mu${\small-}$1p$ dataset. 
The passing fractions are extremely low for either in the $1\mu${\small-}$1p$ sample. Fig.~\ref{fig:PS_14} shows low passing fractions for $\mu^-$ and $p$ and high passing fractions for other three particles in images with the final state of $Ne${\small-}$N\gamma${\small-}$0\mu${\small-}$N\pi${\small-}$0p$.







\begin{figure}[htb!pb]
\centering
\includegraphics[width=8.6cm]{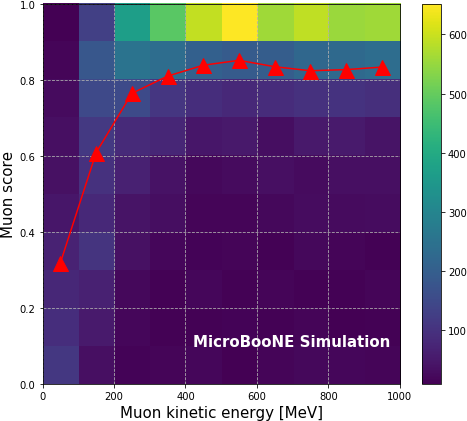}
\vspace{1pt}

\includegraphics[width=8.6cm]{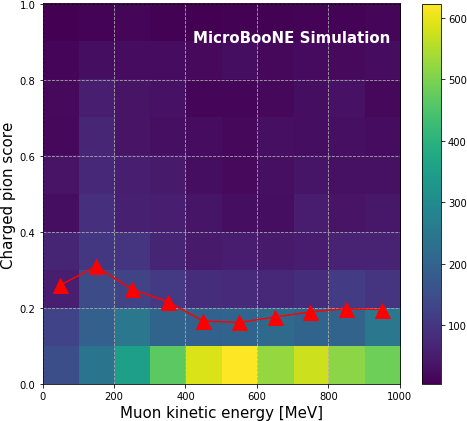}
\caption[]{Muon score vs. muon kinetic energy~(top) and charged pion score vs. muon kinetic energy~(bottom) for the $1\mu${\small-}$1p$ simulation. Red dots indicate the average score in the vertical bin. }
\label{fig:mu_score_mu_eng}
\end{figure}

Figure~\ref{fig:mu_score_mu_eng} shows the correlation between $\mu^-$/$\pi^\pm$ scores and the $\mu^-$ kinetic energy using the same $1\mu${\small-}$1p$ simulation of 10,000 events. One can see that when the $\mu^-$ particles have low kinetic energy and produce fewer $\mu^-$-like pixels in the image, the $\mu^-$ score is decreased. 
Meanwhile, $\pi^\pm$ scores for the same dataset appear to be comparatively low across all tested muon energies.  

\subsection{$1e${\small-}$1p$ Simulated Sample}

\begin{figure}[htb!pb]
\centering
\includegraphics[width=8.6cm]{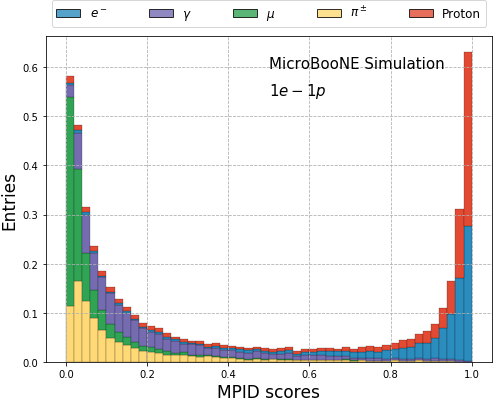}
\caption{~MPID score distributions for the probabilities of {\it p}, $\mu^-$, $e^-$,$\gamma$, $\pi^\pm$ on the $1e${\small-}$1p$  validation sample. \label{fig:mpid-val-1e1p}}
\end{figure}

Figure~\ref{fig:mpid-val-1e1p} shows stacked MPID score distributions for the simulated $1e${\small-}$1p$ dataset. A similar plot for a complementary configuration~($0e${\small-}$N\gamma${\small-}$N\mu${\small-}$N\pi${\small-}$0p$) is shown in Fig.~\ref{fig:PS_9} in appendix~\ref{app1}. MPID correctly calculates high scores for signal particles of {\it p} and $e^-$. One can see between Fig.~\ref{fig:mpid-val-1e1p} and Fig.~\ref{fig:PS_9}, the network shows good separation between track particles in deriving low scores for $\mu^-$ and $\pi^\pm$.  
The MPID CNN also shows good separation between shower-like particles when $e^-$'s are present in the image: derived scores for $\gamma$ are clustered close to zero, while $e^-$-like scores are clustered around unity.  

\begin{figure}[htb!pb]
\centering
\includegraphics[width=8.6cm]{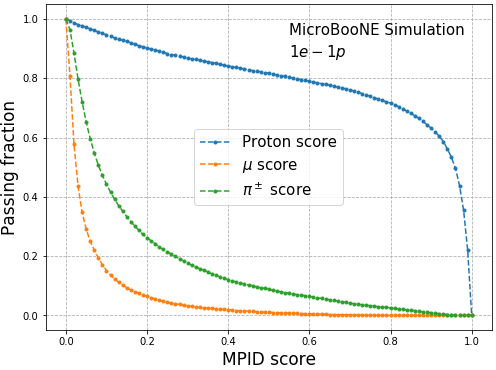}
\vspace{1pt}

\includegraphics[width=8.6cm]{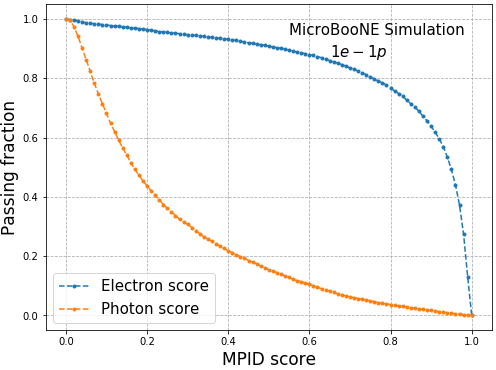}

\caption[]{MPID passing fractions for track-like particles~(top) of {\it p}, $\mu^-$ and $\pi^\pm$ on the $1e${\small-}$1p$ validation sample. MPID passing fractions for shower-like particles~(bottom) of $e^-$ and $\gamma$ on the $1e${\small-}$1p$ validation sample.}
\label{fig:mpid-effi-1e1p}
\end{figure}

The passing fractions over MPID scores for track-like particles in the $1e${\small-}$1p$ dataset are given in Fig.~\ref{fig:mpid-effi-1e1p}. 
Similar plots of the complementary configuration~($0e${\small-}$N\gamma${\small-}$N\mu${\small-}$N\pi${\small-}$0p$) are shown is shown in Fig.~\ref{fig:PS_9} in appendix~\ref{app1}.
The passing fraction for {\it p} in the image are much higher than the fractions for $\mu^-$ or $\pi^\pm$.  
The capability to discriminate between {\it p} and $\mu^-$ appears to be particularly high, while {\it p}/$\pi^\pm$ separation also remains high.  
This difference in performance between $\mu^-$ and $\pi^\pm$ should not be too surprising given the level of $\pi^\pm$-$\mu^-$ passing fractions demonstrated in the previous section.  
Figure~\ref{fig:mpid-effi-1e1p} also shows the passing fractions for the shower-like particles in the $1e${\small-}$1p$ dataset. 
Fig.~\ref{fig:PS_9} shows low passing fractions for $e^-$ and $p$ and high passing fractions for other three particles in images with the final state of $0e${\small-}$N\gamma${\small-}$N\mu${\small-}$N\pi${\small-}$0p$.

\begin{figure}[htb!pb]
\centering
\includegraphics[width=8.6cm]{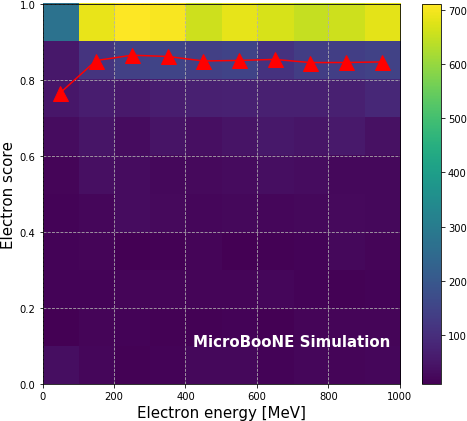}
\vspace{1pt}

\includegraphics[width=8.6cm]{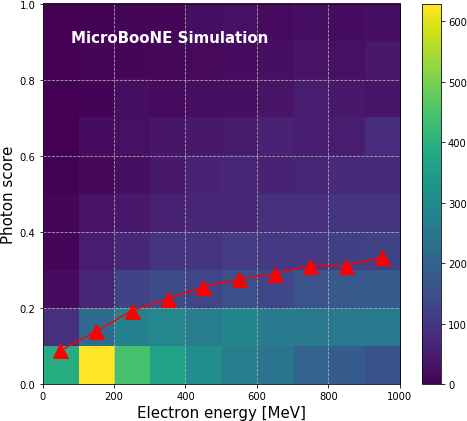}
\caption[]{Electron score vs. electron kinetic energy~(top) and photon score vs. electron kinetic energy~(bottom) for the $1e${\small-}$1p$ simulation. Red dots indicate the average score in the vertical bin.}
\label{fig:e_score_e_eng}
\end{figure}

Figure~\ref{fig:e_score_e_eng} shows the correlation between $e^-$/$\gamma$ scores and $e^-$ kinetic energy.  
One can see the MPID network has an overall high $e^-$ score until the $e^-$ kinetic energy approaches its critical energy in liquid argon and becomes less shower-like.  
In a related sense, $\mu^-$ scores for low energy $1e${\small-}$1p$ interactions are found to be slightly higher than high energy ones.  
Meanwhile, the $\gamma$ score for these events has a positive correlation with $e^-$ kinetic energy, since high energy $e^-$ are more likely to experience substantial amounts of radiative energy loss.  

\section{ Comparison of Data/Simulation Performance}
\label{sec:data}

We prepared two different MicroBooNE LArTPC data samples to validate the performance of the MPID network on data.  
The MPID network was not employed in the selection of these data samples.  
The first data sample is a $1\mu${\small-}$1p$ enriched selection that uses a hybrid selection of a series of reconstruction algorithms~\cite{vtxfinding} and MicroBooNE's semantic segmentation network~\cite{ssnet}. 
This dataset is intended to be used in a MicroBooNE LEE $1e${\small-}$1p$ analysis to provide a data-based constraint on the BNB neutrino beam's intrinsic $\nu_e$ contamination.
The second sample contains $\nu_\mu$ charged current interactions with a final-state $\pi^0$ ($\nu_\mu$CC$\pi^0$) as defined in Ref.~\cite{ub_ccpi0}.
In this section we demonstrate that the MPID network works well on real LArTPC images.  
We show good agreement in MPID scores between data and simulation for the selected datasets.




To enable data/simulation comparisons for these two event classes, we simulate neutrino interactions using the GENIE v3.0.6~\cite{genie} neutrino Monte Carlo generator.  
To accurately include on-surface cosmogenic backgrounds present in all MicroBooNE LArTPC images, beam-off data containing only cosmic rays is overlayed on simulated neutrino interaction images.
Beam-off data is taken with cosmic ray triggers.
An overlay sample is a combination of GENIE simulated beam events and cosmic data events.
This ensures that the reported particle score distributions for data and simulated images will be equally affected by the presence of cosmic rays.  

In the study of the $1\mu${\small-}$1p$ dataset, we apply the MPID network to processed images containing only wire signal activity associated with particles reconstructed at a candidate neutrino interaction vertex.  
In the study of $\nu_\mu$CC$\pi^0$ dataset, we instead apply the MPID network to images made with all pixels near the reconstructed vertex; in this case, particle scores are completely independent of any previous reconstruction. 
We show that the network can purify the desired particle content while maintaining good data-simulation agreement in both the `cleaned'~(input images containing only the reconstructed interactions) and potentially `polluted'~(input images also containing cosmic rays) input images.


\subsection{$1\mu${\small-}$1p$ Enriched Data}
\label{sec:1u1p}

The $1\mu${\small-}$1p$ enriched dataset is selected from a set of MicroBooNE beam-on data corresponding to 4.4 $\times$ 10$^{19}$ protons on target~(POT) in the BNB beam.  
These events consist of exactly two reconstructed particles -- ideally one {\it p} and one $\mu^-$ -- at the candidate interaction vertex. The selection consists of two steps. The first step involves a set of preliminary cuts based on optical information and interaction topology cuts. Candidate $1\mu${\small-}$1p$ interactions are required to have more than a threshold number of photo-electrons recorded in the beam trigger window to be signal. Interaction topology selections require candidates to be located inside the TPC with exactly two fully-contained reconstructed tracks. Topology selections also require an opening angle greater than 0.5~radians. The second step involves two boosted decision trees~(BDT) to make a final $1\mu${\small-}$1p$ selection. The first BDT is trained to separate $1\mu${\small-}$1p$ from the cosmic backgrounds using a simulated $\nu_\mu$ sample and a beam-off cosmic ray only dataset. The second BDT is trained to separate $1\mu${\small-}$1p$ from non-signal neutrino interactions (i.e non-charged current quasi-elastic (CCQE) $\nu_\mu$ interactions, off-vertex $\nu_\mu$ interactions and interactions missing more than 20\% energy in reconstruction) using a simulated $\nu_\mu$ sample. Details of preliminary selection and BDT selections will be documented in detail in future publications. The selection of the dataset described above produces 478 data and 466 simulated input images for processing by the MPID network.  
In the simulated dataset, 94\% of these images contain true neutrino interactions.   
Among these, 314~(67\% of total images) events contain solely one reconstructable final-state {\it p} and $\mu^-$. 

\begin{figure}[htb!pb]
\centering

 \includegraphics[width=8.6cm]{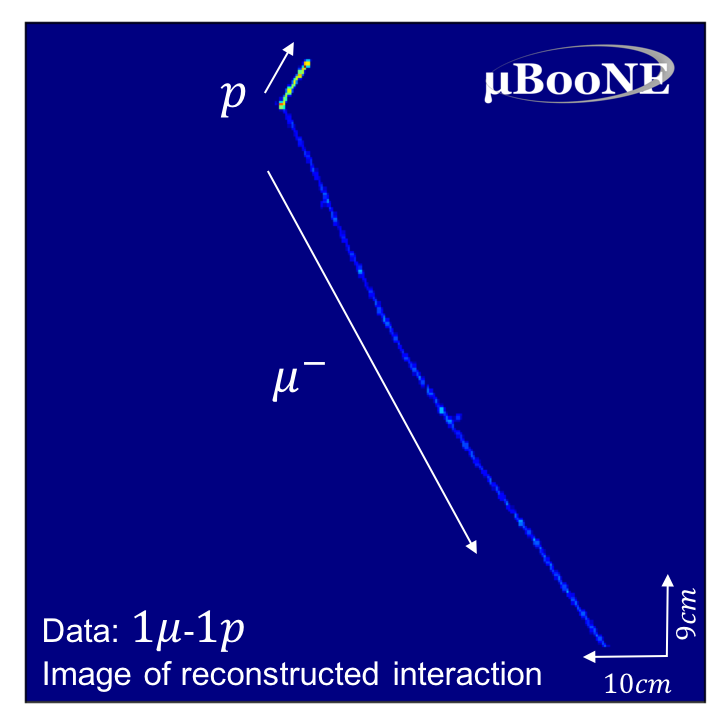}
 
 \caption[]{Example of the input data image from $1\mu${\small-}$1p$ selection. The image is centered at the non-zero pixel weight center. The image has 512$\times$512 pixels. A zoom-in image of 250~$\times$~250 pixels is shown for visualization.}
\label{fig:mpid-1u1p-evd}
\end{figure}

We produce the input images in three steps.  
First, the interaction vertex is located and any associated track-like particles are reconstructed using algorithms described in Ref.~\cite{vtxfinding}; two and only two reconstructed tracks are required.  
Next, a 512$\times$512 image is produced, centered at the pixel-weighted center of the reconstructed $1\mu${\small-}$1p$ event from a flat weight for non-zero pixels. 
Finally, to address noise-related features, a threshold is  placed on the images with a minimum and maximum pixel value of 10 and 500, respectively. 
This procedure removes effects from pixels from unrelated interactions near the neutrino interaction vertex.  
Figure~\ref{fig:mpid-1u1p-evd} shows an example of a $1\mu${\small-}$1p$ image fed into the MPID network. The image is from the collection plane.


\begin{figure}[htb!pb]
\centering

 \includegraphics[width=8.6cm]{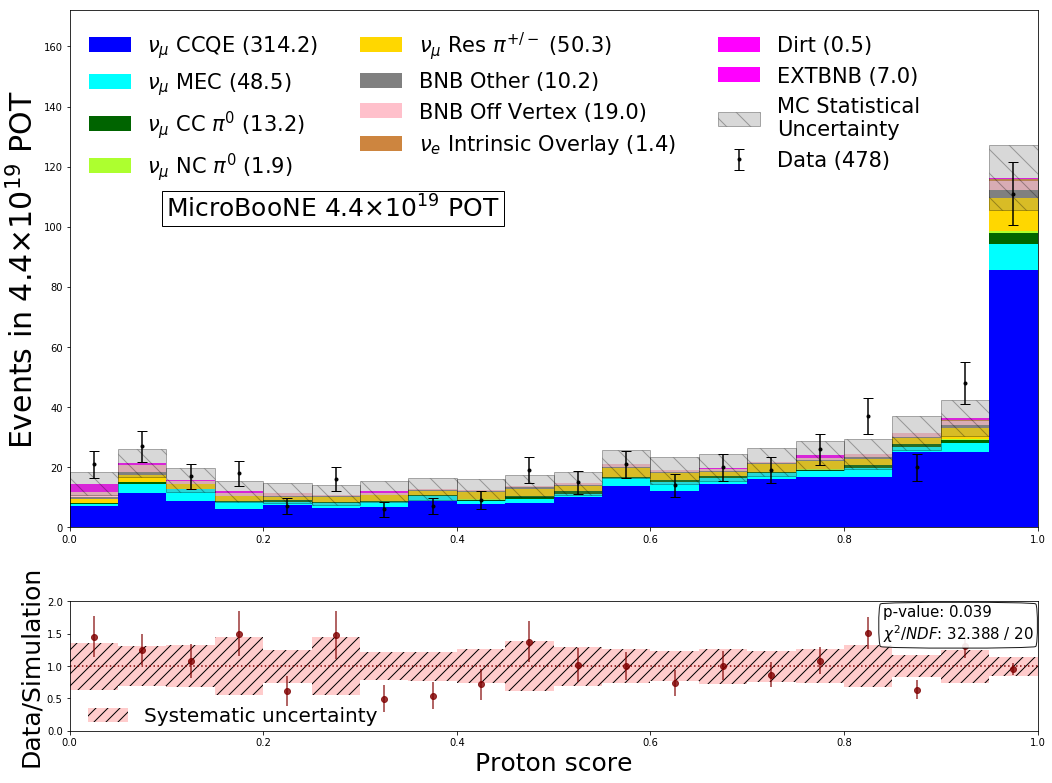}
 \vspace{1pt}

 \includegraphics[width=8.6cm]{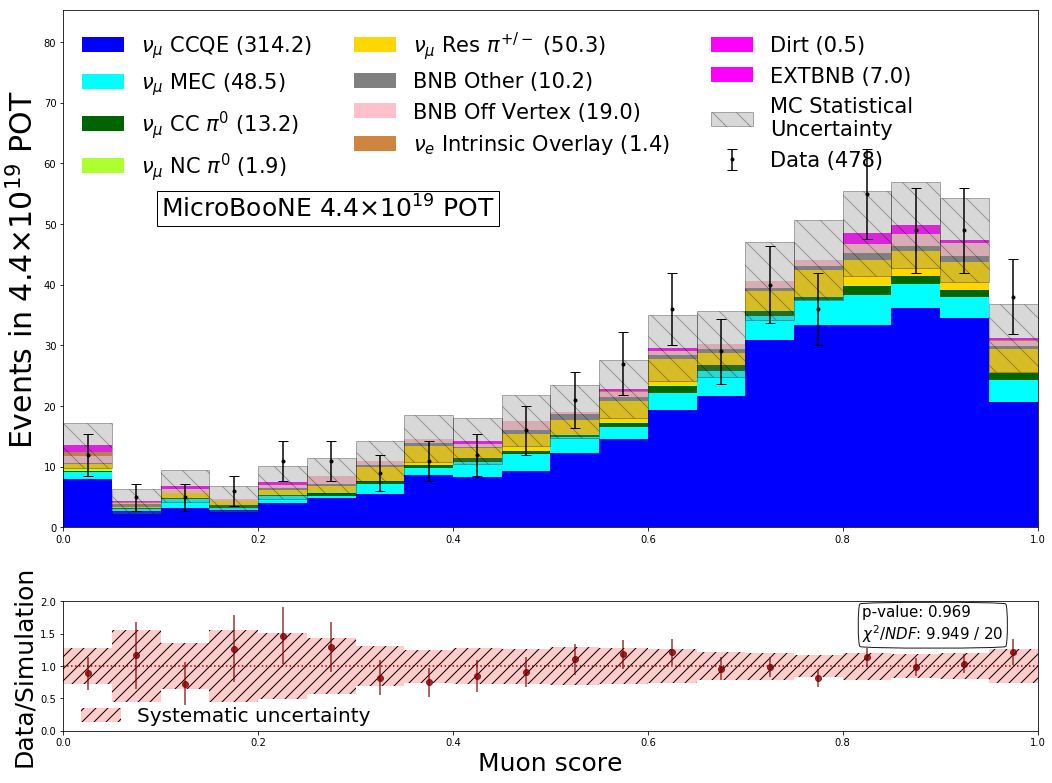}


\caption[]{MPID proton score distribution (top) and muon score distribution (bottom) for selected $1\mu${\small-}$1p$ interactions. Simulation-predicted score distributions show satisfactory agreement with those realized in the $1\mu${\small-}$1p$ selection applied to MicroBooNE data.  Plot error bars indicate data statistical errors, while hatched bands indicate statistical and/or systematic uncertainties in the simulated dataset. The $\chi^2$ calculation incorporates contributions from  systematic and statistical uncertainties. The breakdown of interaction type is based on the predicted event classification for the initial neutrino interaction.}
\label{fig:mpid-1u1p-proton-muon}
\end{figure}

The top image of Fig.~\ref{fig:mpid-1u1p-proton-muon} shows the {\it p} score for the selected candidate $1\mu${\small-}$1p$ interactions, broken down into the true physics process of each imaged vertex.  
The simulation predicts that true $1\mu${\small-}$1p$ charged-current neutrino interactions should cluster at high {\it p} score, with background processes (particularly cosmic processes) more evenly distributed across the score axis.  
In the data, a distinct peak is present at high {\it p} score, providing a strong indication of proton(s) being present in most of the images.  

The bottom sub-panel of this sub-figure shows the ratio of data and simulation versus the {\it p} score.  
We note that as we are primarily concerned with understanding the agreement in the distribution of scores from 0 to 1, 
 discussion of the level of absolute agreement in normalization between data and simulation is beyond the scope of this study.  
For each point, the data's statistical uncertainty is shown, along with the systematic uncertainty associated with flux and cross-section uncertainties.  
Beam flux uncertainties are evaluated by re-weighting events according to the properties of the hadrons that decay to produce the neutrinos.  
Cross section uncertainties are evaluated by re-weighting events according to the properties of the neutrino’s interaction with an argon nucleus. Detector uncertainties are in development and are expected to not have a dominant systematic effect on MPID scores for $1e${\small-}$1p$ events.
Good agreement is found between the data and simulation across the full range of {\it p} scores with flux and cross section uncertainties. 
This level of agreement was quantified by calculating the $\chi^2$ between the data and simulation distributions in the top panel of Fig.~\ref{fig:mpid-1u1p-proton-muon}.  
This $\chi^2$ includes both statistical and systematic uncertainties in the data and simulation.  
A $\chi^2$/NDF of 32.4/~20 is found, indicating a comparable performance of the MPID on both data and simulation.  

Figure~\ref{fig:mpid-1u1p-proton-muon} also shows the $\mu^-$ score distribution for the same selected $1\mu${\small-}$1p$ interactions.   
A majority of events are found in the higher score region, indicating that the MPID algorithm has correctly identified the presence of $\mu^-$ in these images.  
The bottom panel again shows the ratio of data to simulation in the $\mu^-$ score distribution; systematic error bars are similarly defined as for the simulated {\it p} score distribution.  
A $\chi^2$/NDF of 9.9/~20 is found between the two distributions, indicating good MPID data-simulation agreement for $\mu^-$ score.  

\begin{figure}[htb!pb]
\centering
 \includegraphics[width=8.6cm]{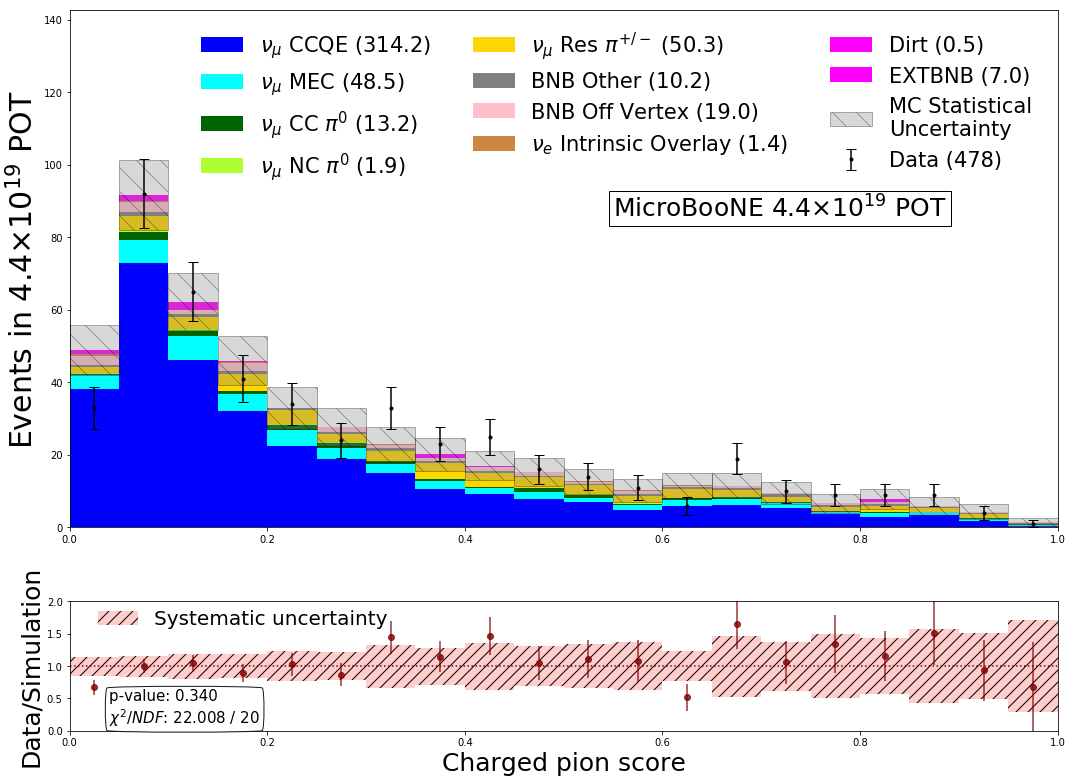}
  \vspace{1pt}
 \includegraphics[width=8.6cm]{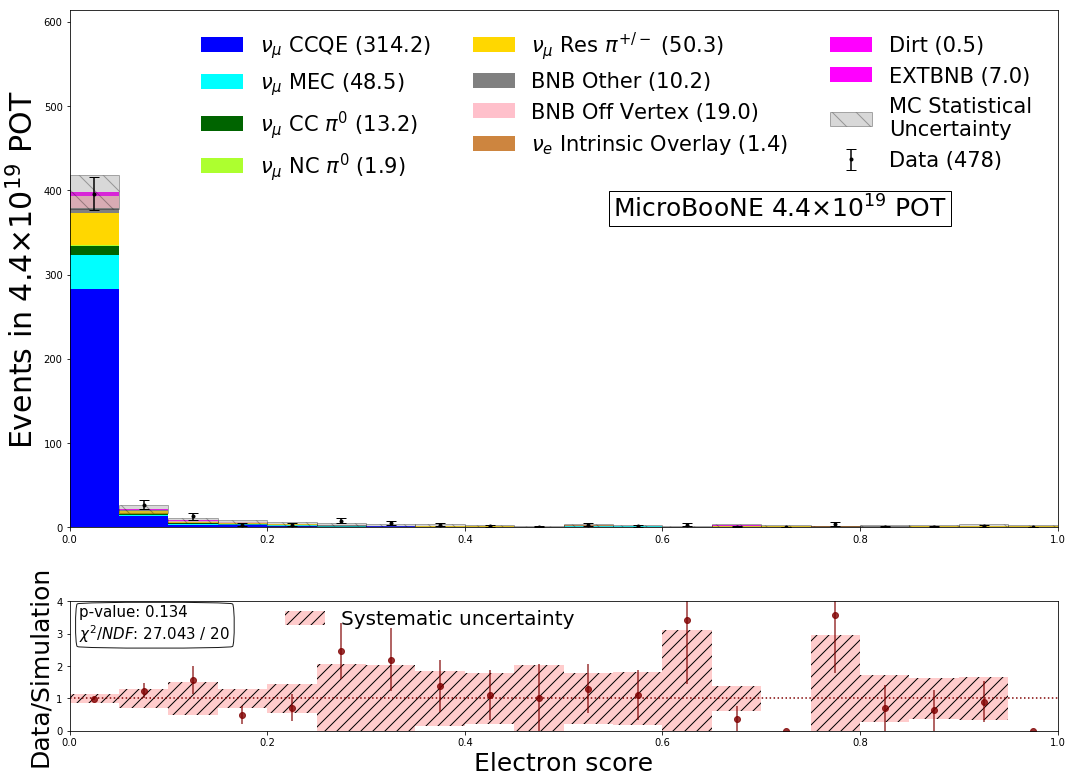}
 \vspace{1pt}
 \includegraphics[width=8.6cm]{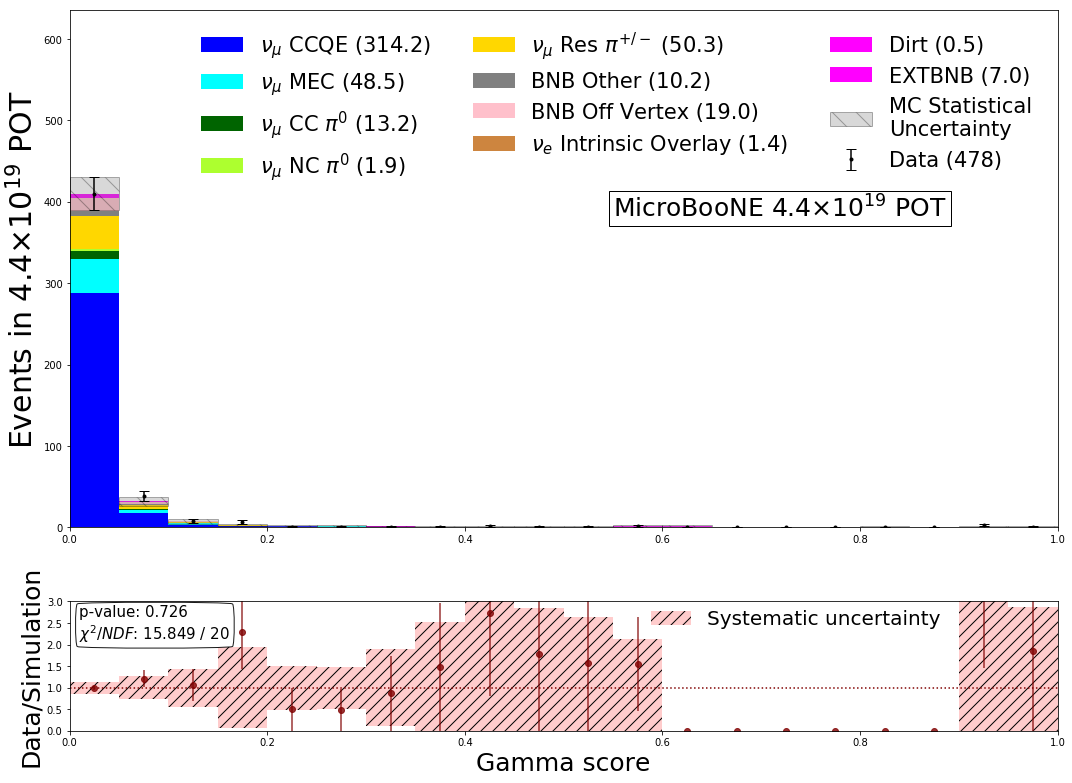}
\caption[]{Charged pion score distribution~(top), electron score distribution~(mid), and photon score distributions~(bottom) for selected $1\mu${\small-}$1p$ interactions. Score distributions agree with the $1\mu${\small-}$1p$ selection. Data and simulation agree well. $\chi^2$ calculation include systematic and statistical uncertainties.}
\label{fig:mpid-1u1p-eminus-gamma-pion}
\end{figure}

Figure~\ref{fig:mpid-1u1p-eminus-gamma-pion} shows the score distributions for particle types expected to be absent from or contained in limited quantities in the selected $1\mu${\small-}$1p$ dataset: $\pi^{\pm}$, $e^-$, and $\gamma$.  
For $\gamma$ and $e^-$, the score distributions are peaked very close to zero, since input images have only track-like particles, and because, as demonstrated in Section~\ref{sec:mc}, discrimination between track-like $\mu^-$ and $p$ particles and shower-like $\gamma$ and $e^-$ particles is expected to be high.  
Scores for track-like $\pi^{\pm}$ particle scores are also clustered towards zero, but with a broader overall width; this result also matches the expectations of Section~\ref{sec:mc}.   
The $\chi^2$/NDF of 22.0/~20, 27.0/~20, and 15.8/~20 for data/simulation comparisons for $\gamma$, $\pi^{\pm}$, and $e^-$ indicate comparable performances of MPID on data and simulation.

The MPID network appears to provide similar performance on both data and simulated neutrino interaction images containing primarily track-like final-state particles.  
This similarity in performance is achieved despite the input image's reliance on other reconstruction algorithms to `remove' pixel content not related to final-state particles connected to the candidate neutrino interaction vertex.  
This indicates that not only the MPID algorithm, but also the upstream reconstruction algorithms, treat data and simulated LArTPC images on an equal footing.

\subsection{$\nu_\mu$CC$\pi^0$ Enriched Data}

A study of $\pi^0$-producing charged current $\nu_\mu$ ($\nu_\mu$CC$\pi^0$) interactions is useful in providing a similar data/simulation agreement validation for images that also contain shower-like objects, as is expected from charged-current $\nu_e$ interactions.  
For this study, we select events from the same dataset used in MicroBooNE's previous $\nu_\mu$CC$\pi^0$ measurement~\cite{ub_ccpi0}.  The primary reconstruction toolkits used to develop selection metrics for these events are Pandora~\cite{pandora} and SSNet~\cite{ssnet}.
Selected events are primarily required to have two showers close to the interaction vertex.   
This requirement makes this dataset distinct from a 1$e-1p$ selection, where one and only one shower is allowed, which must be directly attached to the vertex.  
In this way, in studying MPID performance on the $\nu_\mu$CC$\pi^0$ data sample, we demonstrate not only data/simulation performance, but also show how the network can help to reduce a major intrinsic background to the $\nu_e$ channel: $\pi^0$-producing interactions.  

\begin{figure}[htb!pb]
\centering

\includegraphics[width=8.6cm]{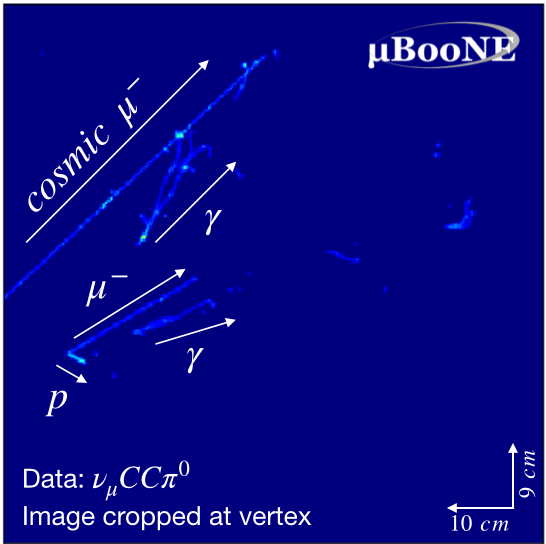}
 
\caption[]{Example of the input data image from the $\nu_\mu$CC$\pi^0$ selection. The image is centered at the reconstructed vertex. The image has $512\times512$ pixels. A zoom-in image of $250\times250$ pixels is shown for visualization.}
\label{fig:mpid-pi0-evd}
\end{figure}

Input images from $\nu_\mu$CC$\pi^0$ candidates are generated by cropping a 512 $\times$ 512 square image centered at the reconstructed interaction vertex, rather than at the image's pixel-weighted center as in the $1\mu${\small-}$1p$ images. 
To ensure that $\pi^0$ decay $\gamma$s are not scrubbed from the image, no additional pixel `cleaning' is applied.  
This means that cosmic rays and other interactions unrelated to the vertex remain in input images, presenting an additional challenge to the MPID network's performance. 
The same noise filtering metric, as described for the $1\mu${\small-}$1p$ dataset, is applied to the images. Figure~\ref{fig:mpid-pi0-evd} shows an example of a $\nu_\mu$CC$\pi^0$-containing image fed into the MPID network. The image is from the collection plane.

The selection and dataset described above produces 2051 data and 2011 simulated input images for processing by the MPID network.  
According to the simulation, 41\% of total events have $\nu_\mu$CC$\pi^0$ interactions and 60\% of events contain $\pi^0$-including interactions~(including the $\nu_\mu$CC$\pi^0$ interactions). 

 \begin{figure}[htb!pb]
\centering
 \includegraphics[width=8.6cm]{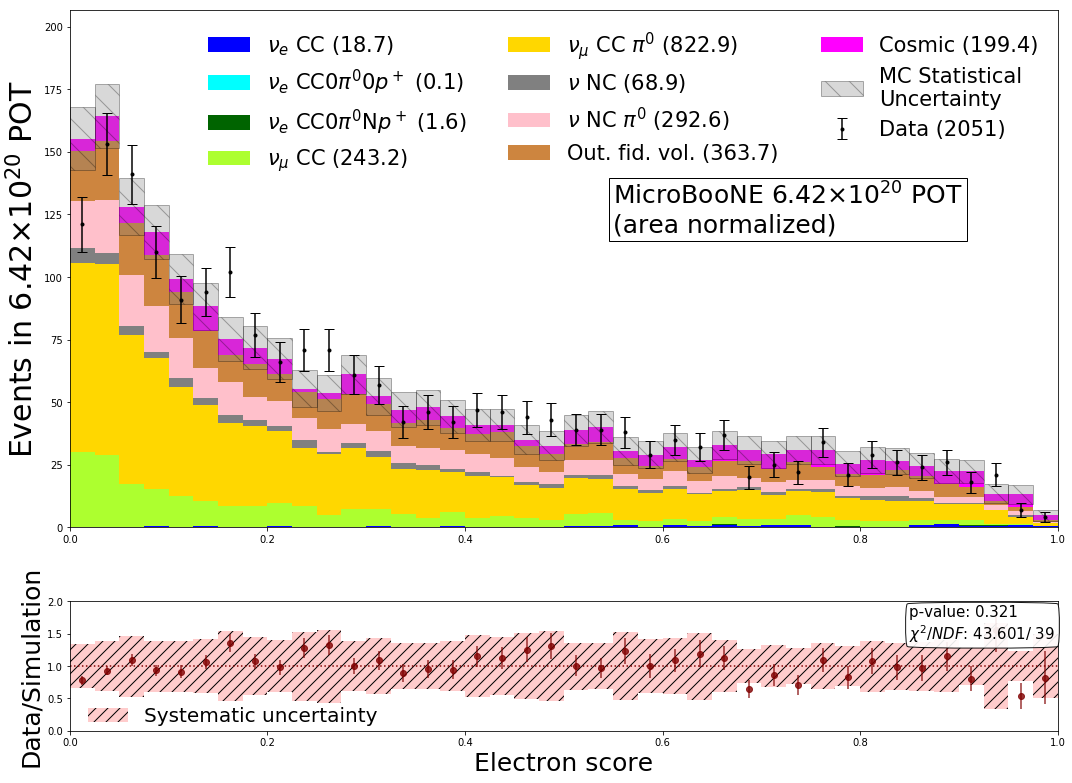}
  \vspace{1pt}

 \includegraphics[width=8.6cm]{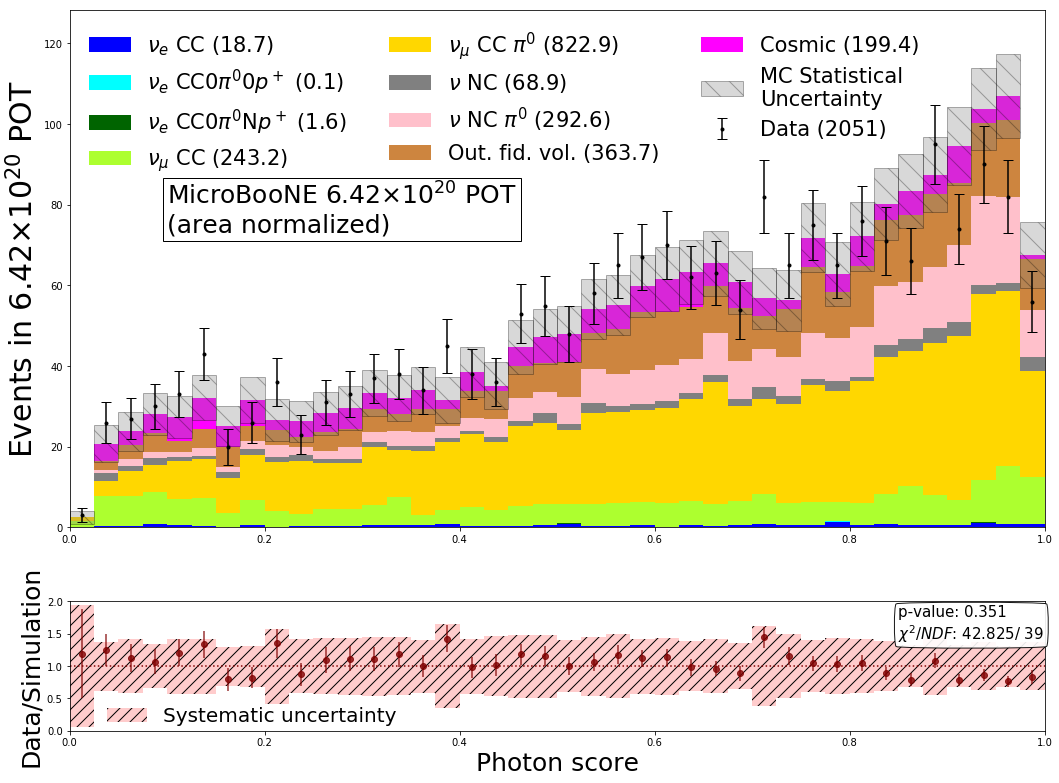}
 \vspace{1pt}



\caption{Electron score distribution (top) and photon score distribution (bottom) for selected $\nu_\mu$CC$\pi^0$ interactions. Score distributions agree with the $\nu_\mu$CC$\pi^0$ selection. Data and simulation agree well. $\chi^2$ calculation include systematic and statistical uncertainties.}
\label{fig:mpid-ccpi0-eminus-gamma}
\end{figure}

\begin{figure}[htb!pb]
\centering
 \includegraphics[width=8.6cm]{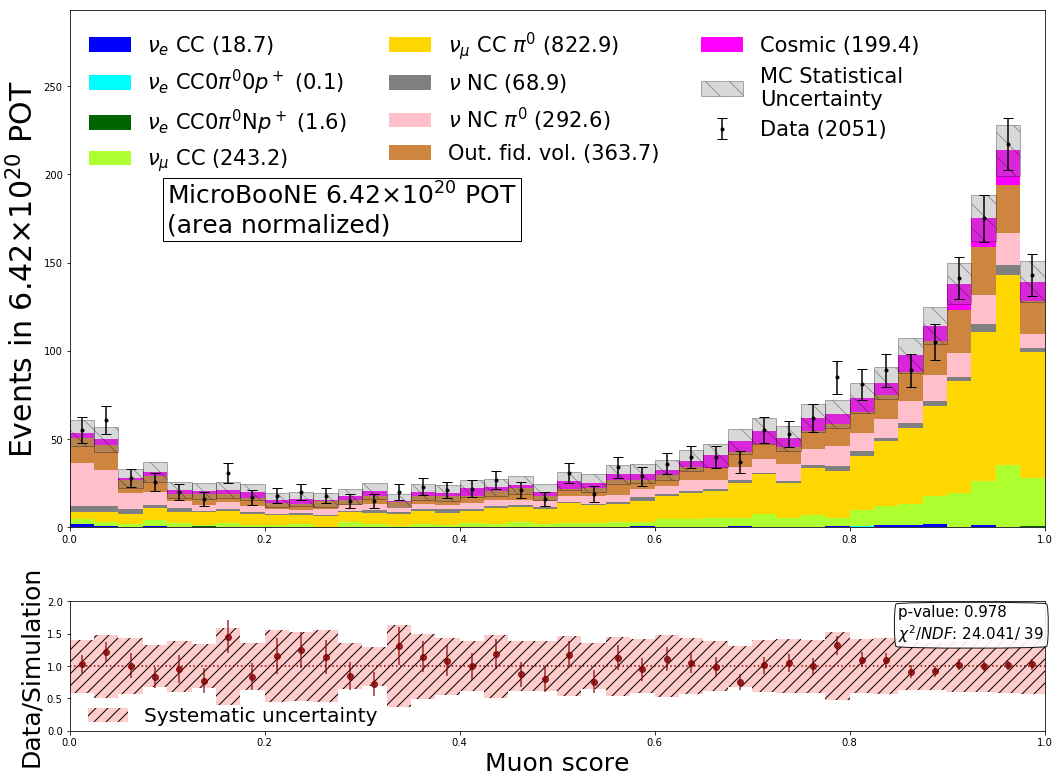}
\caption{
Muon score distribution for selected $\nu_\mu$CC$\pi^0$ interactions. Score distributions for data and simulation agree well using the $\nu_\mu$CC$\pi^0$ selection. $\chi^2$  calculation includes systematic and statistical uncertainties.}
\label{fig:mpid-ccpi0-muon}
\end{figure}

Figure~\ref{fig:mpid-ccpi0-eminus-gamma} shows the score distribution for having any $e^-$ in the images cropped from the $\nu_\mu$CC$\pi^0$ sample.  
The score indicates a generally low probability of having $e^-$-like features in the data and simulated images.  
As a comparison, Fig.~\ref{fig:mpid-ccpi0-eminus-gamma} also shows the score distribution for having any $\gamma$-rays in the images.  
One can see a much higher score distribution for the $\gamma$ existence case, as expected based on the event filtering criteria described above.  
Figure~\ref{fig:mpid-ccpi0-muon} shows the score distribution for having any $\mu^-$ in the $\nu_\mu$CC$\pi^0$ images. The score generally indicates a high probability of having $\mu^-$-like features in data and simulation. 
In particular, it shows a difference between the CC-and NC-$\pi^0$ events in the low $\mu^-$ score region.

The bottom panels of Fig.~\ref{fig:mpid-ccpi0-eminus-gamma} and Fig.~\ref{fig:mpid-ccpi0-muon} show the ratio of data to simulation versus $e^-$, $\gamma$ and $\mu^-$ scores following a area-only comparison.  
Systematic uncertainties are also included in the same manner as described for the 1$\mu$-1$p$ dataset.  
Good comparable performance can be seen between data and simulation with $\chi^2$/NDFs of 43.6/~39 for $e^-$ score,  42.8/~39 for $\gamma$ score and 24.0/~39 for $\mu^-$ score.  
Thus, this study demonstrates that, for a subset of $\pi^0$-containing neutrino interactions, the MPID algorithm can reliably identify shower-related particle content in images without introducing biases between neutrino data and simulation predictions. This is achieved despite the presence of additional incidental pixel activity being present in interaction candidate images.

\section{Use of MPID In A Low Energy Excess Measurement}
\label{sec:nue_mc}

In the two previous sections, we have demonstrated the MPID network's utility in particle identification for both track and shower topologies in LArTPC images, as well as its equivalent performance on both data and simulated events.  
We will now apply the trained MPID network to simulated BNB $\nu_e$ and $\nu_{\mu}$ interactions overlayed with beam-off cosmic event images to demonstrate the ability of the MPID network to aid in event selection for MicroBooNE's deep learning-based $1e${\small-}$1p$ low-energy excess search.

\subsection{Simulated Intrinsic $\nu_e$ vs. $\nu_\mu$CCQE and $\nu_\mu\pi^0$}


We generated simulated neutrino events to evaluate the performance of MPID in the $1e${\small-}$1p$ selection in identifying beam-intrinsic backgrounds originating from from $\nu_{\mu}$CCQE and neutrino interactions with one or more $\pi^0$s in the final state~($\nu_\mu\pi^0$).
Samples for these three datasets are produced using the standard GENIE v3.0.6~\cite{genie} neutrino interaction generator and filtered using truth-level information.  
In these samples, we require the lepton kinetic energy be greater than 35 MeV and {\it p} kinetic energy greater than 60 MeV. The minimum kinetic energy thresholds were set in order to choose events whose lepton and {\it p} trajectories are long enough to be reconstructed by our deep learning based vertex finding and particle reconstruction algorithms~\cite{vtxfinding}.
Samples are then processed using the reconstruction algorithms to identify candidate interaction vertices and nearby related particles.  
Finally, input images are generated with pixels from only the reconstructed interaction final-state particles; each interaction is required to have two particles at this stage. Images are centered at the pixel weighted center of reconstructed interactions.
No other selection cuts beyond the truth-level filtration described above are applied to the samples.  

\begin{figure}[htb!pb]
\centering
\includegraphics[width=8.6cm]{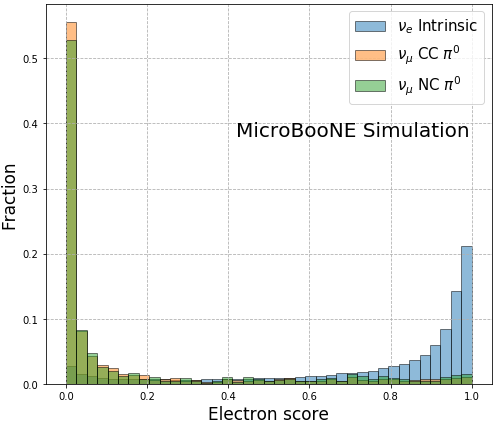}
\caption{Electron score of $\nu_e$ intrinsic events and $\nu_\mu\pi^0$ events.
Both datasets are generated with the GENIE neutrino generator and filtered using truth level information. Presented events have a reconstructed vertex.\label{fig:mc_eminus_nue_pi}}
\end{figure}

Figure~\ref{fig:mc_eminus_nue_pi} shows the $e^-$ score distribution of reconstructed events from $\nu_e$ and $\nu_\mu\pi^0$ datasets.  
A good separation is visible between these two event classes.  
For example, with only an $e^-$ cut score of 0.5, 83\% of $\nu_\mu$NC$\pi^0$ and 86\% of $\nu_\mu$CC$\pi^0$ events are rejected, while 81\% of true $1e${\small-}$1p$ events are selected.
It seems likely that further gains in background rejection could be achieved by also considering scores for other particles and by using differing input pixel image inclusion settings.  

Previous discussion from the occlusion analysis presented in Section~\ref{sec:mc} provides some level of insight into the causes of the substantial discrimination shown in Fig.~\ref{fig:mc_eminus_nue_pi}.  
In particular, $\nu_e$ interactions will contain a shower-like object with a trunk directly connected to another particle, a feature that was clearly noticed by the MPID network.  
This is not the case for most $\gamma$ rays present in $\nu_\mu\pi^0$ interactions.  
Another critical parameter for separating $e^-$- and $\pi^0$-including events is the energy deposition density, dE/dx, along this vertex-connected shower trunk; the trunk region information is usually well-reconstructed, since it is almost always directly attached to neutrino candidate vertex. 
Some of the discrimination in Fig.~\ref{fig:mc_eminus_nue_pi} may thus also arise from the network's ability to discriminate a high trunk dE/dx for vertex-connected showers from quickly-converting $\pi^0$ $\gamma$ rays.  

\begin{figure}[htb!pb]
\centering
\includegraphics[width=8.6cm]{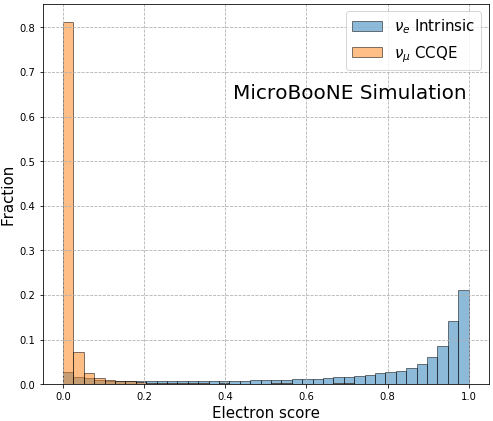}
\caption{Electron score between $\nu_e$ intrinsic events and $\nu_\mu$ CCQE events. 
Both datasets are generated with the GENIE neutrino generator and filtered using truth level information. Presented events have a reconstructed vertex.\label{fig:mc_eminus_nue_ccqe}}
\end{figure}

The $e^-$ score can also be applied to separate $1e${\small-}$1p$ and $1\mu${\small-}$1p$ events. The separation is shown in Fig.~\ref{fig:mc_eminus_nue_ccqe}. 
The $\nu_e$ and $\nu_{\mu}$CCQE events are well separated using the $e^-$ score calculated by the MPID network.  
For example, with only an $e^-$ cut score of 0.2, 91\% of true $1e${\small-}$1p$ events are selected, while 95\% of $\nu_{\mu}$CCQE events are rejected.
This discrimination ability almost certainly arises from the lack of shower-like topologies in the $\nu_{\mu}$CCQE interaction images.  

\subsection{Simulated Intrinsic $\nu_e$ vs. Cosmic Event}


Due to the lack of substantial overburden and the long readout time, cosmic rays could provide a substantial background to a BNB-based $1e${\small-}$1p$ $\nu_e$ measurement in MicroBooNE.  
As most of this cosmic ray activity is induced by $\mu^-$, it is expected that the presence of a {\it p} in the signal's final state will aid in distinguishing the two categories.  
To test the MPID network's ability to discriminate the signal's {\it p} particle content, 
we generated a simulated intrinsic $\nu_e$ dataset with cosmic data overlay, in addition to another event set consisting purely of beam-off cosmic triggers.  
For both datasets, we applied the vertex finding and particle reconstruction algorithms developed for two-track events, as described in Ref.~\cite{vtxfinding}; in particular, each image is required to have exactly two reconstructed particles connected to the candidate neutrino interaction vertex. 
As in the sub-section above, no selection cuts are applied beyond truth-level event filtration. 

\begin{figure}[htb!pb]
\centering
\includegraphics[width=8.6cm]{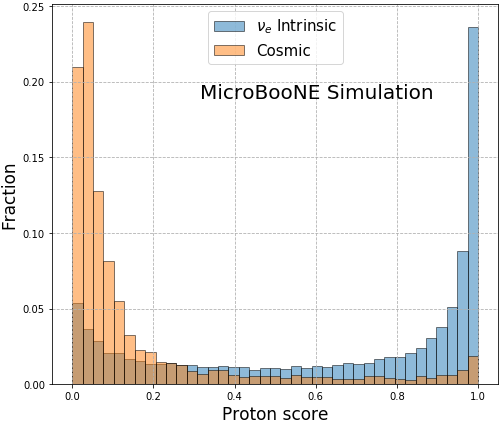}
\caption{Proton score of $\nu_e$ intrinsic events and beam-off cosmic data. 
The $\nu_e$ dataset is generated with the GENIE neutrino generator and filtered using truth level information. Presented events have a reconstructed vertex. \label{fig:mc_proton_nue_cosmic}}
\end{figure}

Figure~\ref{fig:mc_proton_nue_cosmic} shows the {\it p} score distributions on images from the intrinsic $\nu_e$ dataset with cosmic overlay and the pure cosmics dataset. 
One can see that the majority of pure cosmic dataset events reconstructed as two-particle signals events have {\it p} scores below 0.2.  
Meanwhile, the majority of reconstructed $\nu_e$ intrinsic events have {\it p} scores near 1. For example, with only a {\it p} cut score of 0.5, 81\% of true $1e${\small-}$1p$ events are selected, while 79\% of cosmic events are rejected.

Investigation of information from prior reconstruction stages and hand-scanning of event displays indicates that the small peak in {\it p} score close to zero in the $\nu_e$ intrinsic dataset is due to inefficiencies in {\it p} reconstruction as shown in Fig. 21(a) of~Ref.~\cite{vtxfinding}.  
Similar investigations show that the small peak of of {\it p} score close to one in the cosmic sample is introduced by cosmics with small incident angles relative to the collection plane; these non-{\it p} tracks are often topologically compressed by reconstruction algorithms, giving them the appearance of short tracks with a proton-like Bragg peak.  
Thus, future improvements in lower-level signal processing and particle reconstruction is likely to further improve the cosmic discrimination shown in Fig.~\ref{fig:mc_proton_nue_cosmic}.

\section{Conclusion}

We have developed a CNN-based multiple particle identification network, MPID, and applied it to images of event interactions in MicroBooNE data.  
This is the first demonstration of the performance of a CNN that incorporates systematic uncertainties in LArTPC data, and the first use of CNNs to perform particle identification on real LArTPC data.
The network takes a 512$\times$512 LArTPC image and calculates the probability scores for any particle in the image as {\it p}, $e^-$, $\gamma$, $\mu^-$, and $\pi^\pm$. The training images are generated with a customized event generator that concatenates particles at the same vertex. The code for making the network and training sample are made available in MPID~\cite{mpid_github} and LArSoft~\cite{uboonecode}.

10,000 $1e${\small-}$1p$ and $1\mu${\small-}$1p$ images are used to benchmark the network performance on simulated interactions. Pass fractions of particles present in the images are found to surpass those not present in the input images.



Satisfactory agreement in all score distributions are found 
between data and simulation despite the many complexities of the MicroBooNE liquid argon TPC response, including inactive wire regions~\cite{elec_noise}, electronics noise~\cite{elec_noise}, signal processing~\cite{sp_1, sp_2}, and space charge effects~\cite{spacecharge}.

We also demonstrated the metrics and performance of applying the MPID network on BNB beam data from MicroBooNE, which also illustrated the MPID network's clear capabilities in particle discrimination.  
When we take reconstructed vertex activity as input in filtered $1\mu${\small-}$1p$ candidate event images, MPID score distributions are indeed high for $p$ and $\mu^-$, and low for $e^-$, $\gamma$ and $\pi^\pm$.  
When we instead take all pixel activity as input in filtered images containing $\pi^0$-produced $\gamma$ rays, we see large differences between obtained $e^-$ and $\gamma$ scores.  
By applying these demonstrated particle identification capabilities to simulated BNB $\nu_e$ and $\nu_{\mu}$ interactions,  we have shown that this validated tool can play an important role in achieving a successful low-energy electron-like excess measurement in MicroBooNE.

\acknowledgments

This document was prepared by the MicroBooNE collaboration using the
resources of the Fermi National Accelerator Laboratory (Fermilab), a
U.S. Department of Energy, Office of Science, HEP User Facility.
Fermilab is managed by Fermi Research Alliance, LLC (FRA), acting
under Contract No. DE-AC02-07CH11359.  MicroBooNE is supported by the
following: the U.S. Department of Energy, Office of Science, Offices
of High Energy Physics and Nuclear Physics; the U.S. National Science
Foundation; the Swiss National Science Foundation; the Science and
Technology Facilities Council (STFC), part of the United Kingdom Research and Innovation;
 and The Royal Society (United Kingdom).  Additional support for the laser
calibration system and cosmic ray tagger was provided by the Albert
Einstein Center for Fundamental Physics, Bern, Switzerland.

\appendix

\section{MPID scores for all particle combinations}
\label{app1}

\begin{figure*}[htb!pb]
\centering
\includegraphics[width=8.6cm]{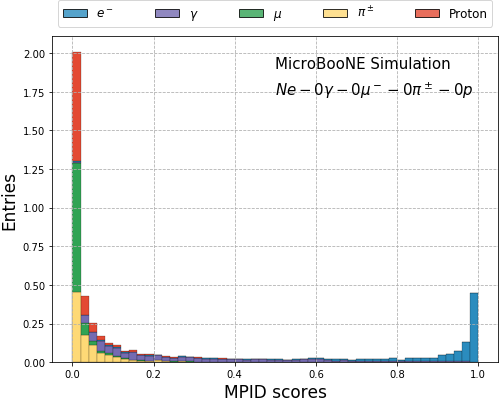}
\includegraphics[width=8.6cm]{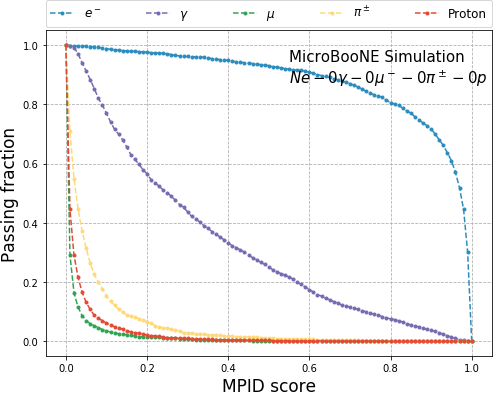}
  \vspace{1pt}
\includegraphics[width=8.6cm]{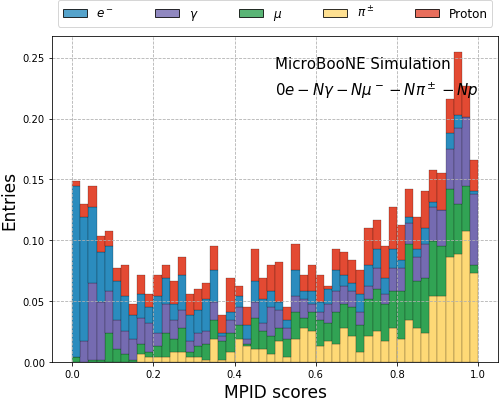}
\includegraphics[width=8.6cm]{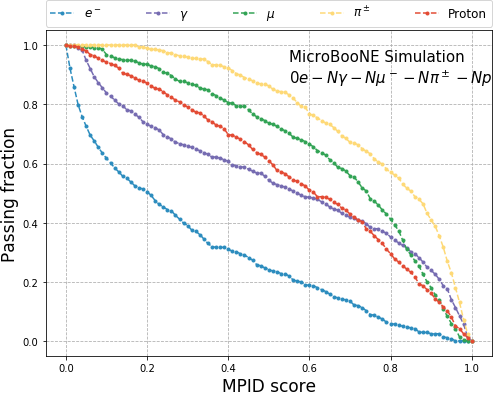}
\caption[]{~MPID score distributions and MPID passing fractions on a complementary set of $Ne${\small-}$0\gamma${\small-}$0\mu${\small-}$0\pi${\small-}$0p$ and $0e${\small-}$N\gamma${\small-}$N\mu${\small-}$N\pi${\small-}$Np$. N is randomly one or two in each event. }
\label{fig:PS_1}
\end{figure*}

This section serves to supplement~Section~\ref{sec:mc} in providing a complete description of the performance of the MPID network on a variety of simulated final-state particle combinations.  
In this section we present the network performances on the full set of different samples over all possible final particle state particle combinations.  
There are 32 different combinations regarding the five considered particle types.  
However, cases involving none of the particle types, as well as all five particle types, were not included in the training or test samples.  
The remaining 30 combinations are presented in this paper.  
For each combination we present a stacked distribution similar to Fig.~\ref{fig:mpid-val-1mu1p} and a passing fraction plot similar to Fig.~\ref{fig:mpid-effi-1mu1p} for each of the five type of particles.  
We present the 30 combinations in 15 pairs, with each pair having two complementary configurations, for example the network performances over the final states of $Ne-0\gamma-0\mu^--0\pi^\pm-0p$ and $0e-N\gamma-N\mu^--N\pi^\pm-Np$ as shown in Fig.~\ref{fig:PS_1}. The data is generated using the same configuration for the test sample described in Section~\ref{training_sample}.  
For 80\% of the sample, particles are simulated with kinetic energies between 60~MeV and 400~MeV for protons and between 30~MeV and 1~GeV for other particles.  
For the other 20\% of the sample, particles are simulated with kinetic energies between 40~MeV and 100~MeV for protons and between 30~MeV and 100~MeV for other particles. Particles are generated with a flat energy distribution.  

\begin{figure*}[htb!pb]
\centering
\includegraphics[width=8.6cm]{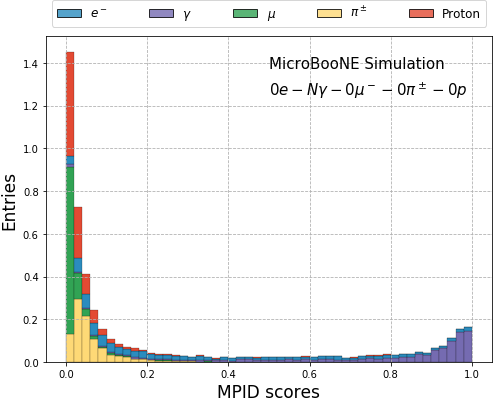}
\includegraphics[width=8.6cm]{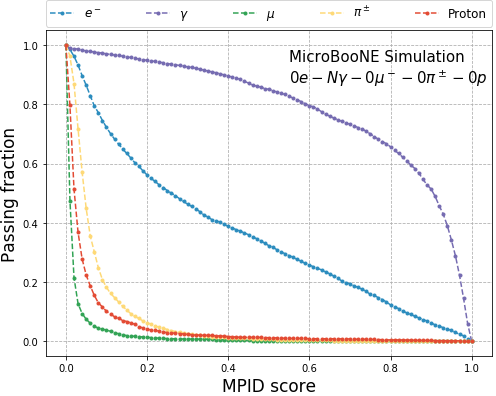}
  \vspace{1pt}
\includegraphics[width=8.6cm]{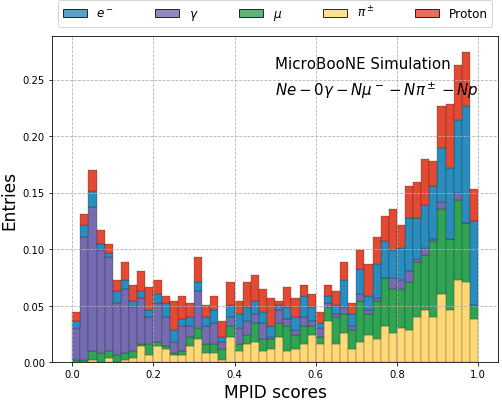}
\includegraphics[width=8.6cm]{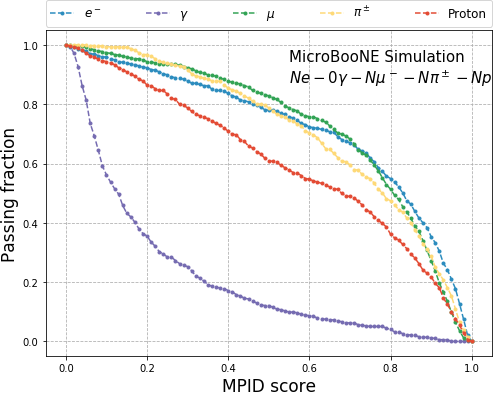}
\caption[]{~MPID score distributions and MPID passing fractions on a complementary set of $0e${\small-}$N\gamma${\small-}$0\mu${\small-}$0\pi${\small-}$0p$ and $Ne${\small-}$0\gamma${\small-}$N\mu${\small-}$N\pi${\small-}$Np$. N is randomly one or two in each event. }
\label{fig:PS_2}
\end{figure*}

\begin{figure*}[htb!pb]
\centering
\includegraphics[width=8.6cm]{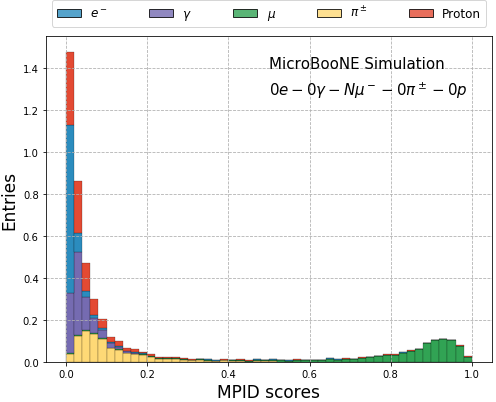}
\includegraphics[width=8.6cm]{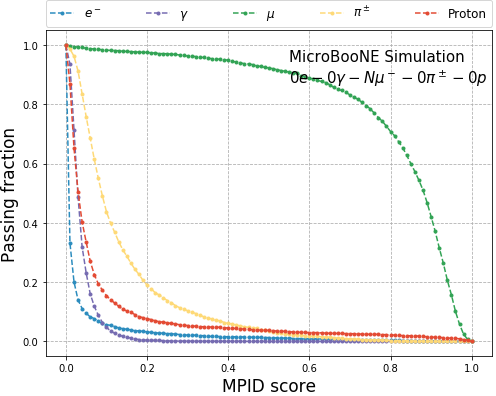}
  \vspace{1pt}
\includegraphics[width=8.6cm]{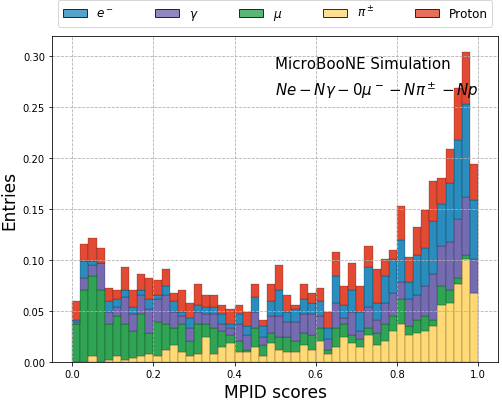}
\includegraphics[width=8.6cm]{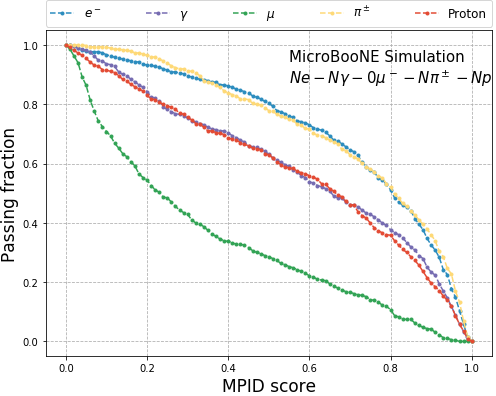}
\caption[]{~MPID score distributions and MPID passing fractions on a complementary set of $0e${\small-}$0\gamma${\small-}$N\mu${\small-}$0\pi${\small-}$0p$ and $Ne${\small-}$N\gamma${\small-}$0\mu${\small-}$N\pi${\small-}$Np$. N is randomly one or two in each event. }
\label{fig:PS_3}
\end{figure*}

\begin{figure*}[htb!pb]
\centering
\includegraphics[width=8.6cm]{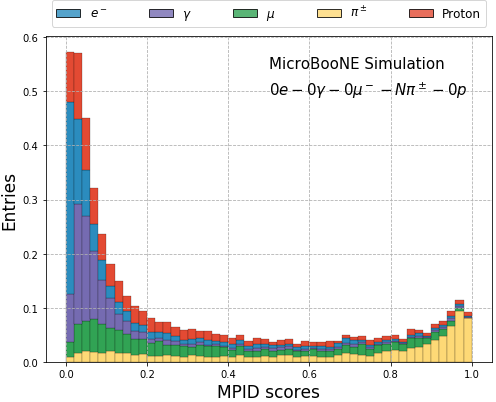}
\includegraphics[width=8.6cm]{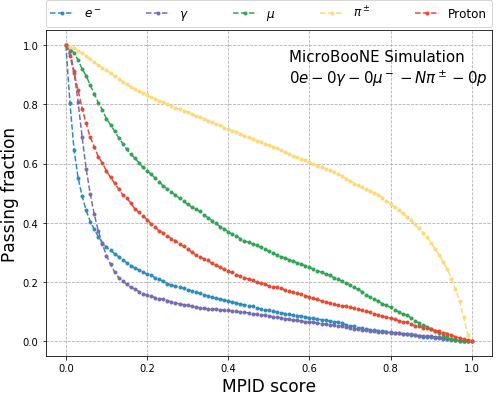}
  \vspace{1pt}
\includegraphics[width=8.6cm]{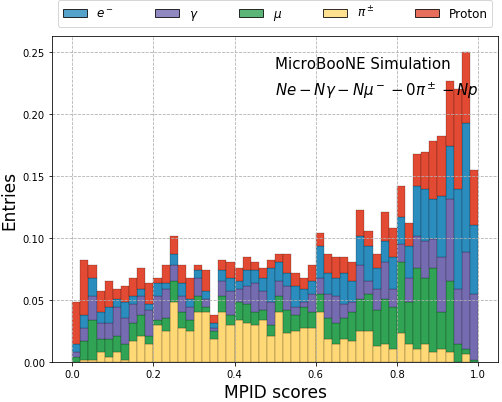}
\includegraphics[width=8.6cm]{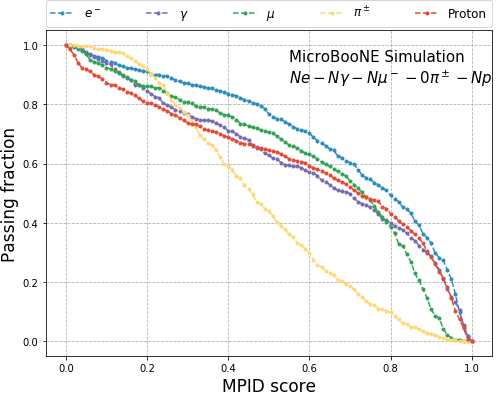}
\caption[]{~MPID score distributions and MPID passing fractions on a complementary set of $0e${\small-}$0\gamma${\small-}$0\mu${\small-}$N\pi${\small-}$0p$ and $Ne${\small-}$N\gamma${\small-}$N\mu${\small-}$0\pi${\small-}$Np$. N is randomly one or two in each event. }
\label{fig:PS_4}
\end{figure*}

\begin{figure*}[htb!pb]
\centering
\includegraphics[width=8.6cm]{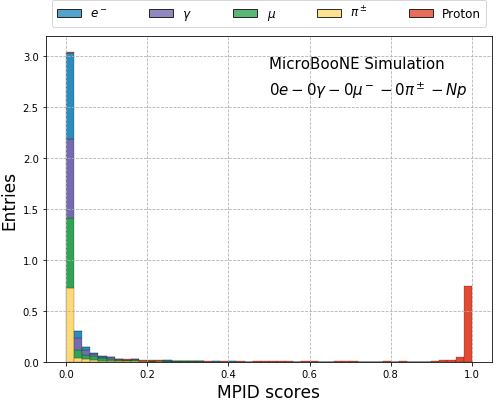}
\includegraphics[width=8.6cm]{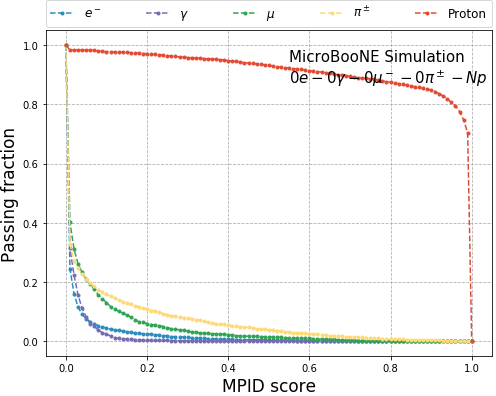}
  \vspace{1pt}
\includegraphics[width=8.6cm]{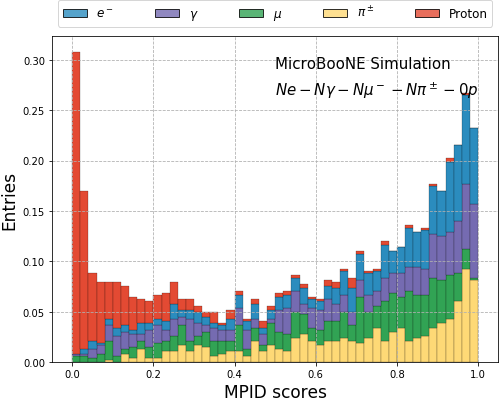}
\includegraphics[width=8.6cm]{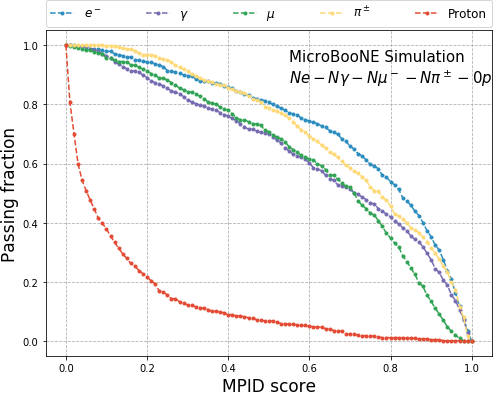}
\caption[]{~MPID score distributions and MPID passing fractions on a complementary set of $0e${\small-}$0\gamma${\small-}$0\mu${\small-}$0\pi${\small-}$Np$ and $Ne${\small-}$N\gamma${\small-}$N\mu${\small-}$N\pi${\small-}$0p$. N is randomly one or two in each event. }
\label{fig:PS_5}
\end{figure*}

\begin{figure*}[htb!pb]
\centering
\includegraphics[width=8.6cm]{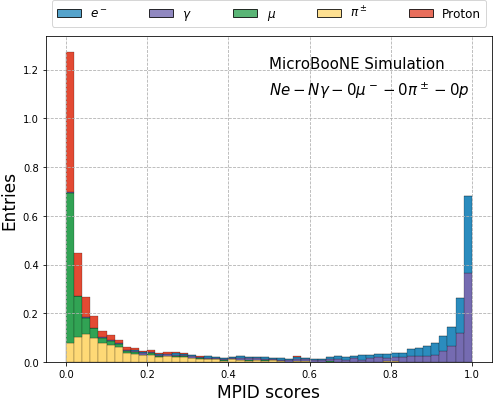}
\includegraphics[width=8.6cm]{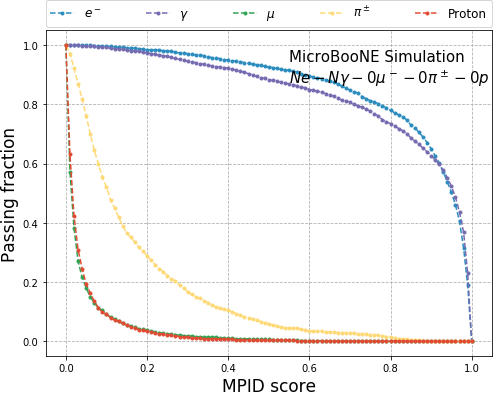}
  \vspace{1pt}
\includegraphics[width=8.6cm]{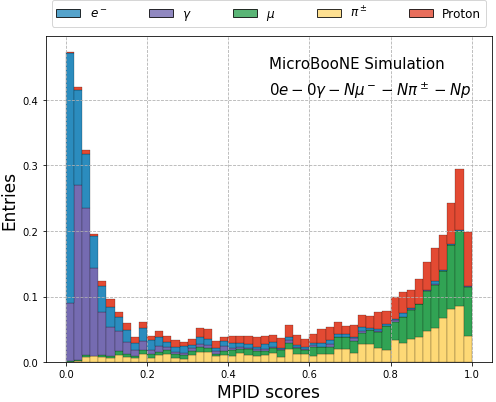}
\includegraphics[width=8.6cm]{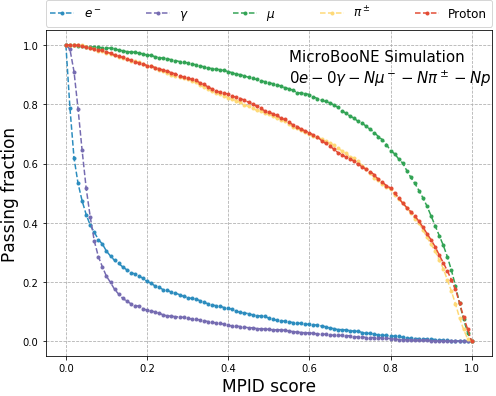}
\caption[]{~MPID score distributions and MPID passing fractions on a complementary set of $Ne${\small-}$N\gamma${\small-}$0\mu${\small-}$0\pi${\small-}$0p$ and $0e${\small-}$0\gamma${\small-}$N\mu${\small-}$N\pi${\small-}$Np$. N is randomly one or two in each event. }
\label{fig:PS_6}
\end{figure*}

\begin{figure*}[htb!pb]
\centering
\includegraphics[width=8.6cm]{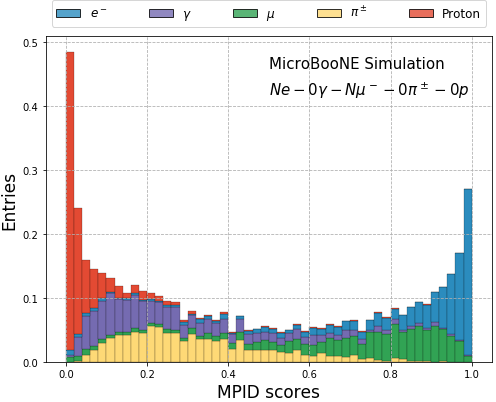}
\includegraphics[width=8.6cm]{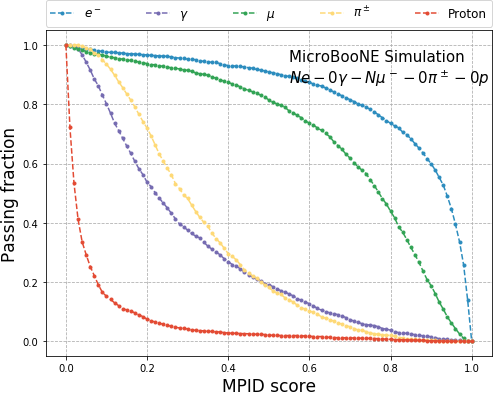}
  \vspace{1pt}
\includegraphics[width=8.6cm]{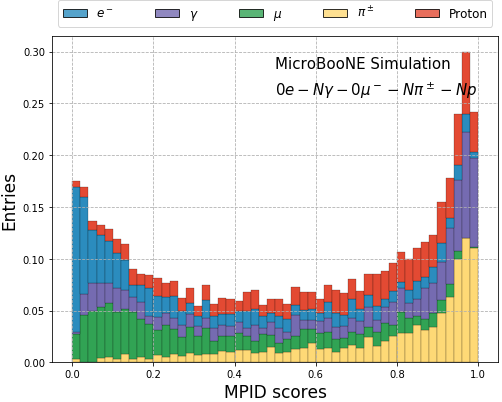}
\includegraphics[width=8.6cm]{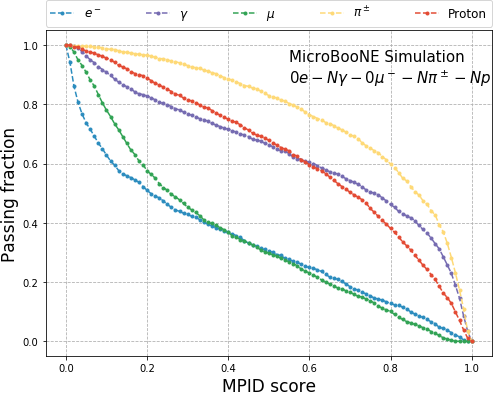}
\caption[]{~MPID score distributions and MPID passing fractions on a complementary set of $Ne${\small-}$0\gamma${\small-}$N\mu${\small-}$0\pi${\small-}$0p$ and $0e${\small-}$N\gamma${\small-}$0\mu${\small-}$N\pi${\small-}$Np$. N is randomly one or two in each event. }
\label{fig:PS_7}
\end{figure*}

\begin{figure*}[htb!pb]
\centering
\includegraphics[width=8.6cm]{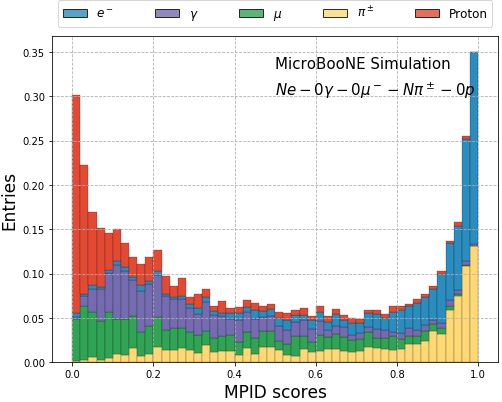}
\includegraphics[width=8.6cm]{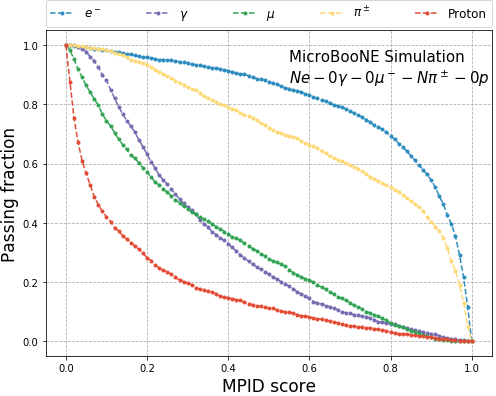}
  \vspace{1pt}
\includegraphics[width=8.6cm]{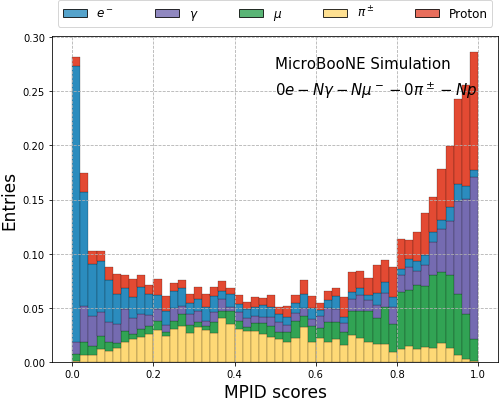}
\includegraphics[width=8.6cm]{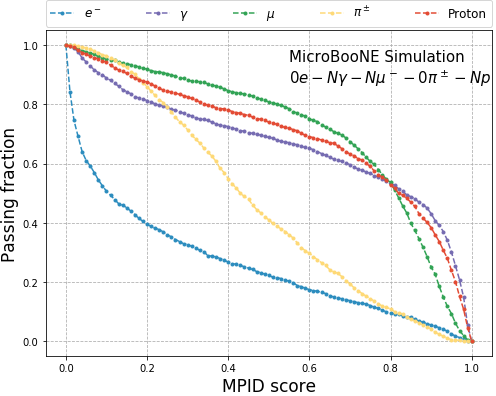}
\caption[]{~MPID score distributions and MPID passing fractions on a complementary set of $Ne${\small-}$0\gamma${\small-}$0\mu${\small-}$N\pi${\small-}$0p$ and $0e${\small-}$N\gamma${\small-}$N\mu${\small-}$0\pi${\small-}$Np$. N is randomly one or two in each event. }
\label{fig:PS_8}
\end{figure*}

\begin{figure*}[htb!pb]
\centering
\includegraphics[width=8.6cm]{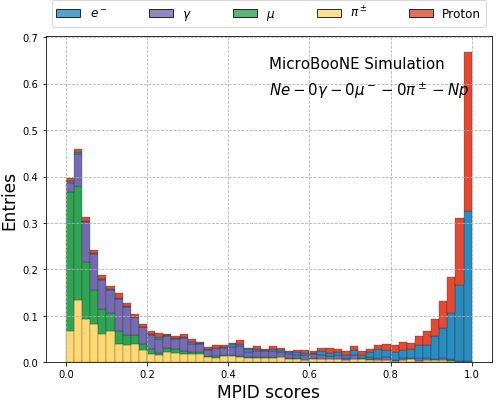}
\includegraphics[width=8.6cm]{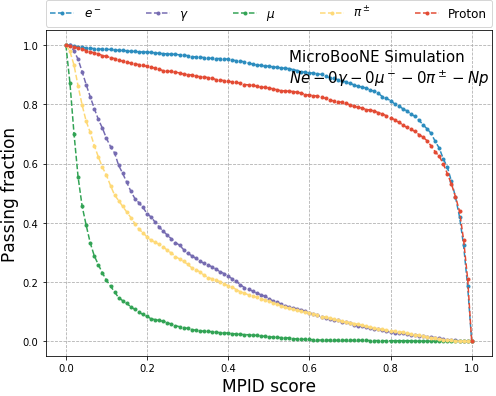}
  \vspace{1pt}
\includegraphics[width=8.6cm]{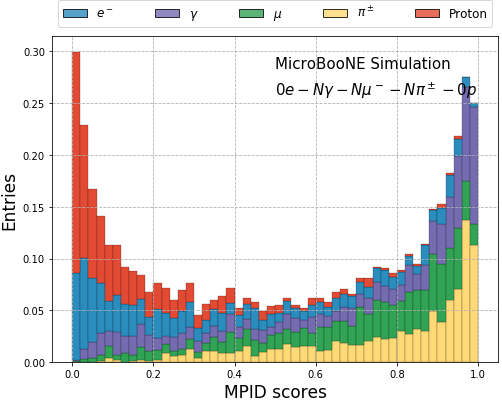}
\includegraphics[width=8.6cm]{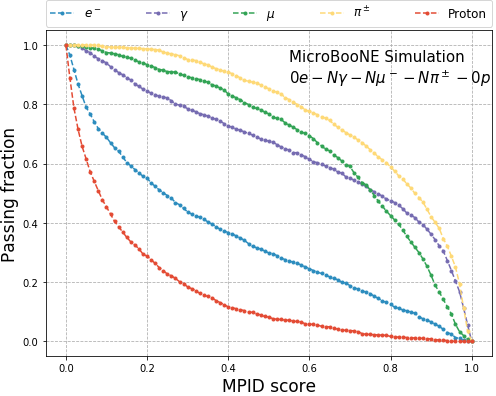}
\caption[]{~MPID score distributions and MPID passing fractions on a complementary set of $Ne${\small-}$0\gamma${\small-}$0\mu${\small-}$0\pi${\small-}$Np$ and $0e${\small-}$N\gamma${\small-}$N\mu${\small-}$N\pi${\small-}$0p$. N is randomly one or two in each event. }
\label{fig:PS_9}
\end{figure*}

\begin{figure*}[htb!pb]
\centering
\includegraphics[width=8.6cm]{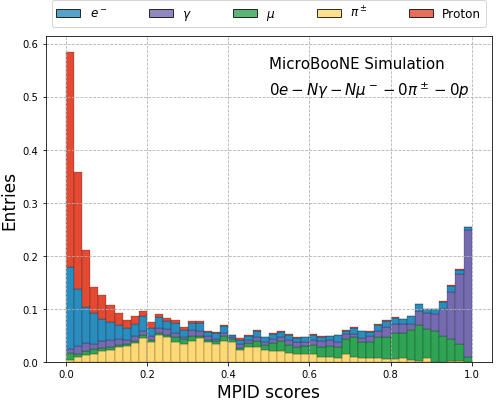}
\includegraphics[width=8.6cm]{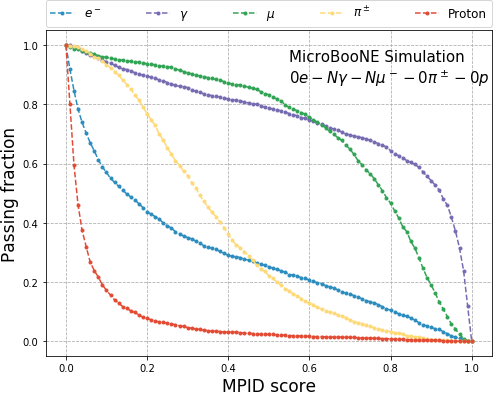}
  \vspace{1pt}
\includegraphics[width=8.6cm]{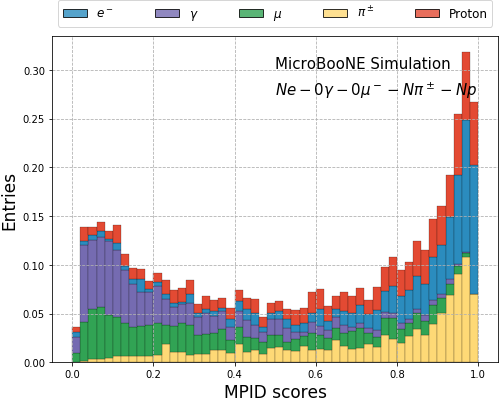}
\includegraphics[width=8.6cm]{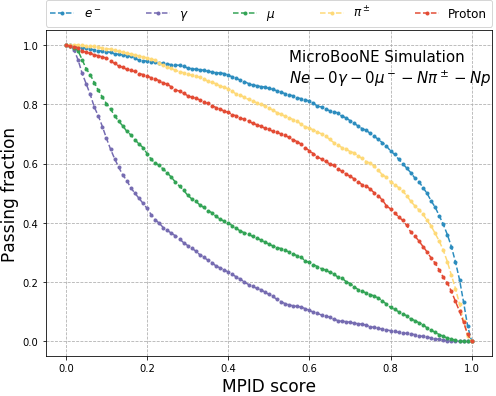}
\caption[]{~MPID score distributions and MPID passing fractions on a complementary set of $0e${\small-}$N\gamma${\small-}$N\mu${\small-}$0\pi${\small-}$0p$ and $Ne${\small-}$0\gamma${\small-}$0\mu${\small-}$N\pi${\small-}$Np$. N is randomly one or two in each event. }
\label{fig:PS_10}
\end{figure*}

\begin{figure*}[htb!pb]
\centering
\includegraphics[width=8.6cm]{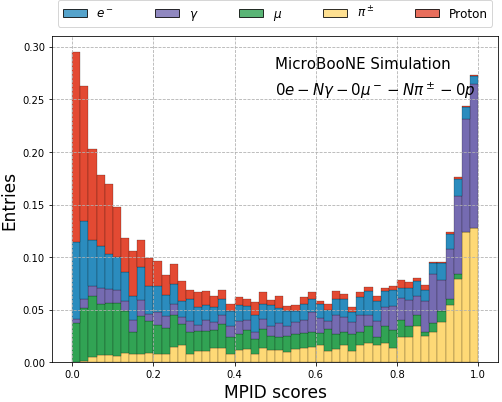}
\includegraphics[width=8.6cm]{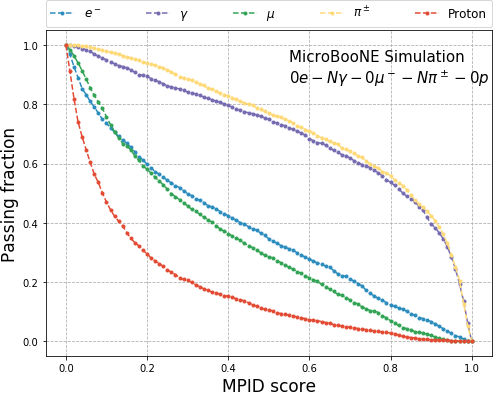}
  \vspace{1pt}
\includegraphics[width=8.6cm]{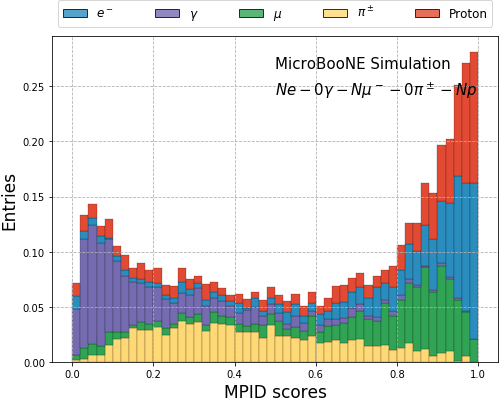}
\includegraphics[width=8.6cm]{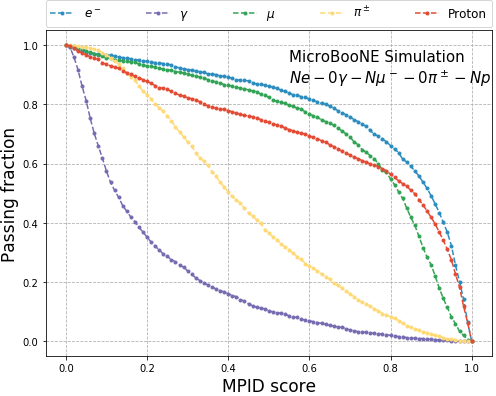}
\caption[]{~MPID score distributions and MPID passing fractions on a complementary set of $0e${\small-}$N\gamma${\small-}$0\mu${\small-}$N\pi${\small-}$0p$ and $Ne${\small-}$0\gamma${\small-}$N\mu${\small-}$0\pi${\small-}$Np$. N is randomly one or two in each event. }
\label{fig:PS_11}
\end{figure*}

\begin{figure*}[htb!pb]
\centering
\includegraphics[width=8.6cm]{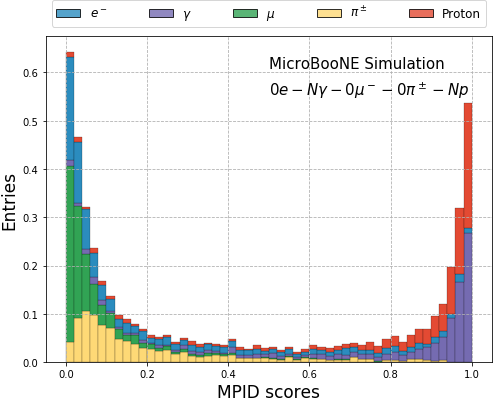}
\includegraphics[width=8.6cm]{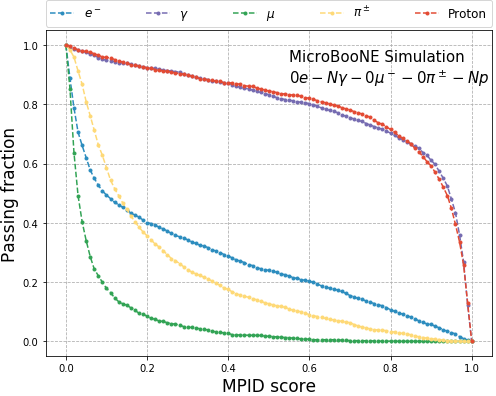}
  \vspace{1pt}
\includegraphics[width=8.6cm]{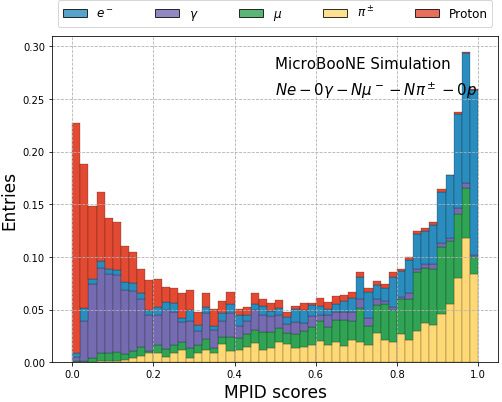}
\includegraphics[width=8.6cm]{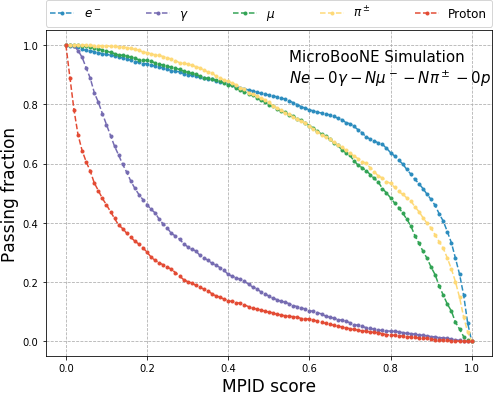}
\caption[]{~MPID score distributions and MPID passing fractions on a complementary set of $0e${\small-}$N\gamma${\small-}$0\mu${\small-}$0\pi${\small-}$Np$ and $Ne${\small-}$0\gamma${\small-}$N\mu${\small-}$N\pi${\small-}$0p$. N is randomly one or two in each event. }
\label{fig:PS_12}
\end{figure*}

\begin{figure*}[htb!pb]
\centering
\includegraphics[width=8.6cm]{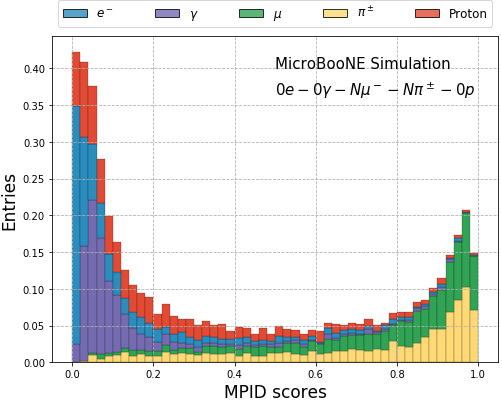}
\includegraphics[width=8.6cm]{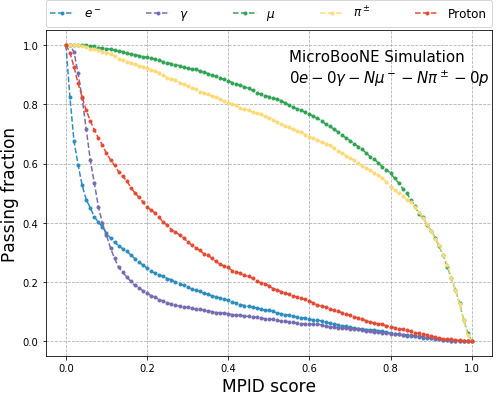}
  \vspace{1pt}
\includegraphics[width=8.6cm]{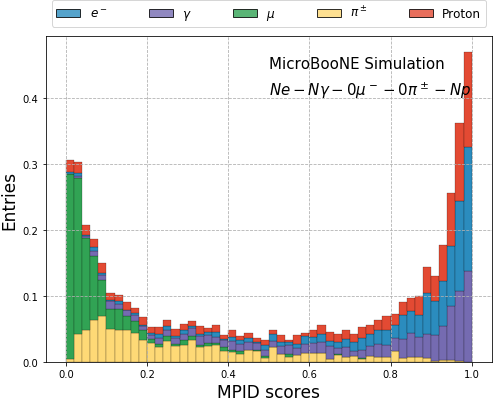}
\includegraphics[width=8.6cm]{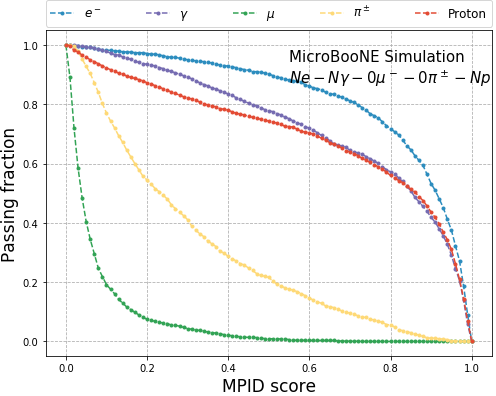}
\caption[]{~MPID score distributions and MPID passing fractions on a complementary set of $0e${\small-}$0\gamma${\small-}$N\mu${\small-}$N\pi${\small-}$0p$ and $Ne${\small-}$N\gamma${\small-}$0\mu${\small-}$0\pi${\small-}$Np$. N is randomly one or two in each event. }
\label{fig:PS_13}
\end{figure*}

\begin{figure*}[htb!pb]
\centering
\includegraphics[width=8.6cm]{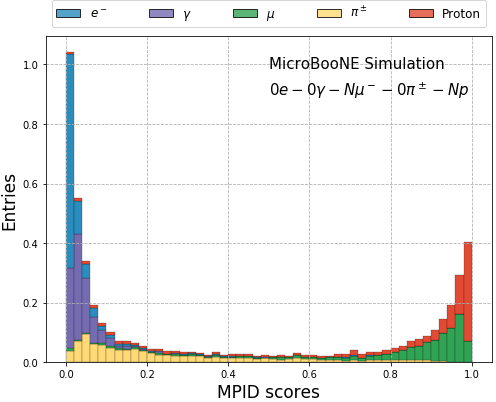}
\includegraphics[width=8.6cm]{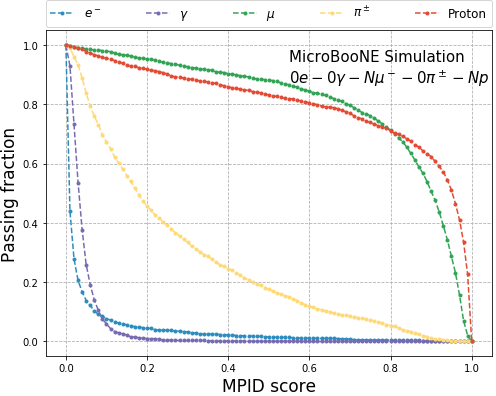}
  \vspace{1pt}
\includegraphics[width=8.6cm]{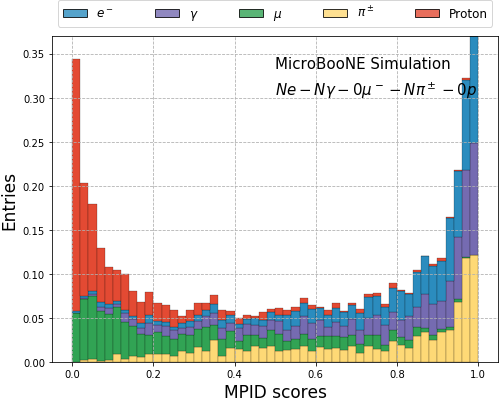}
\includegraphics[width=8.6cm]{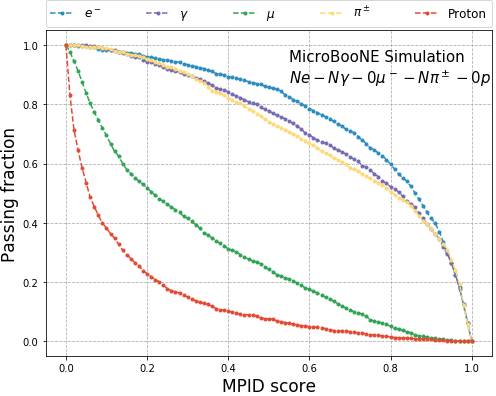}
\caption[]{~MPID score distributions and MPID passing fractions on a complementary set of $0e${\small-}$0\gamma${\small-}$N\mu${\small-}$0\pi${\small-}$Np$ and $Ne${\small-}$N\gamma${\small-}$0\mu${\small-}$N\pi${\small-}$0p$. N is randomly one or two in each event. }
\label{fig:PS_14}
\end{figure*}

\begin{figure*}[htb!pb]
\centering
\includegraphics[width=8.6cm]{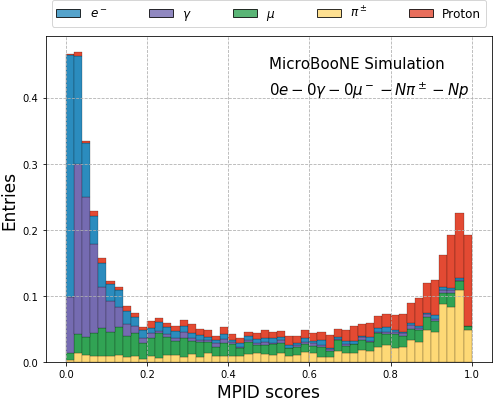}
\includegraphics[width=8.6cm]{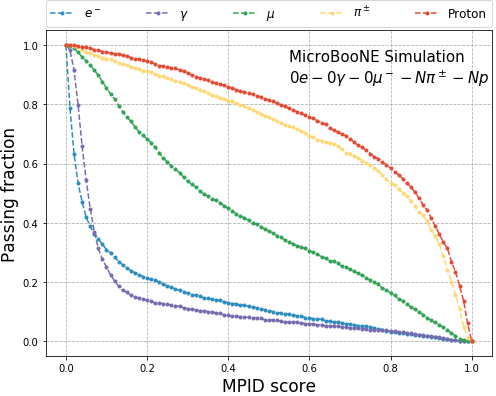}
  \vspace{1pt}
\includegraphics[width=8.6cm]{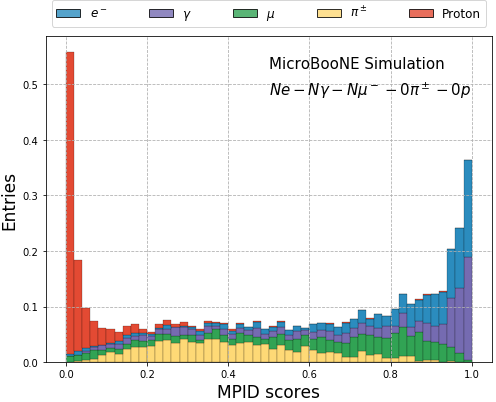}
\includegraphics[width=8.6cm]{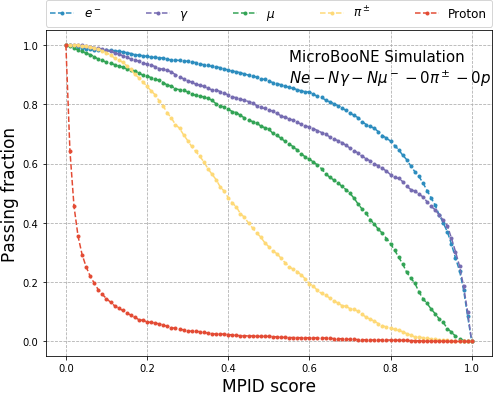}
\caption[]{~MPID score distributions and MPID passing fractions on a complementary set of $0e${\small-}$0\gamma${\small-}$0\mu${\small-}$N\pi${\small-}$Np$ and $Ne${\small-}$N\gamma${\small-}$N\mu${\small-}$0\pi${\small-}$0p$. N is randomly one or two in each event. }
\label{fig:PS_15}
\end{figure*}

\bibliography{reference}

\end{document}